\PassOptionsToPackage{natbibapa}{apacite}
\documentclass[draft]{agujournal2019}

\usepackage{url} 
\usepackage[inline]{trackchanges} 
\usepackage{soul}
\usepackage{xcolor}
\usepackage{xspace}
\usepackage{booktabs}

\usepackage{subcaption}
\usepackage{rotating}
\usepackage{amssymb}
\usepackage{graphicx}
\usepackage{float}
\usepackage{booktabs}
\usepackage[section]{placeins}
\usepackage{tabularx, array}
\usepackage{xurl}
\usepackage{hyperref}
\usepackage{amsmath}
\newcommand{\perkins}{\citet{perkins2026}}

\newcommand{\degree}{$^\circ$\xspace}
\newcommand{\ace}{ACE2S-X-SHiELD\xspace}
\newcommand{\hiroace}{HiRO-ACEv2\xspace}

\draftfalse

\journalname{JGR: Machine Learning and Computation}

\begin{document}

%
%


\title{
    Extreme events and statistics in an AI emulator of a global storm-resolving model
}

%
%




\authors{
Anna Kwa\affil{1}, W. Andre Perkins\affil{1}, Troy Arcomano\affil{1},  Noah D. Brenowitz\affil{3}, Christopher S. Bretherton\affil{1}, Spencer K. Clark\affil{1,2}, Lucas M. Harris\affil{2}, Jeremy McGibbon\affil{1}, Michael S. Pritchard\affil{3}, Oliver Watt-Meyer\affil{1} 
}

\affiliation{1}{Allen Institute for Artificial Intelligence, Seattle, WA, USA}
\affiliation{2}{NOAA/Geophysical Fluid Dynamics Laboratory, Princeton, NJ, USA}
\affiliation{3}{NVIDIA, Santa Clara, CA USA}




\correspondingauthor{Anna Kwa}{annak@allenai.org}



\begin{keypoints}
\item The HiROv2 downscaling model generates realistic precipitation, 10-meter horizontal winds, and sea level pressure for extreme events at 3~km resolution.
\item 10 year climatologies of 3~km resolution \hiroace outputs are similar to the reference X-SHiELD distribution up to the 99.9999th percentile values.
\item Free-running \hiroace simulations generate extreme events (e.g. tropical cyclones, strong downslope winds) with small-scale structure and intensities similar to the reference dataset.
\end{keypoints}

\begin{abstract}

We present \hiroace (High Resolution Output Ai2 Climate Emulator), a two-stage framework for AI emulation of the 3~km resolution X-SHiELD global storm-resolving model (GSRM), and highlight its ability to generate large ensembles of data for studying the characteristics of tropical cyclones and extreme events in high-resolution simulations at an inference computational cost over 300 times lower than that of traditional numerical GSRMs.
\hiroace expands upon the precipitation-only HiRO-ACEv1 model to generate coherent precipitation, 10~m winds, and sea level pressure outputs, enabling a broader range of events to be simulated at high-resolution.
Fine-resolution predictions are generated by running an autoregressive emulator of coarsened X-SHiELD at 100 km, whose multivariate outputs are regionally downscaled to 3~km resolution via a separately trained diffusion model. 
Particular importance is placed on capturing the statistics of extreme events during the diffusion downscaling step. 
Our ability to match the tails of distributions and generate realistic-looking extreme events in both the `perfect prediction' case as well as in free-running emulator rollouts is improved via the use of multiple denoisers at inference time and several updates to the data sampling, noise sampling, and loss weighting during training.
\hiroace closely matches the high-resolution reference distributions of precipitation and near-surface winds out to their 99.9999th percentile values, and broadly tracks the reference distribution of tropical cyclone intensities.

\end{abstract}

\section*{Plain Language Summary}
Numerical models capable of simulating Earth's atmosphere at spatial scales of $\sim$kilometers are useful for studying extreme events across the globe such as tropical cyclones and wind storms.
However, they are computationally expensive, which makes it hard to run simulations over the long time ranges ($\gtrsim$10 years) needed to gather sufficient data for studying these rare occurrences.
In this work we present HiRO-ACEv2, an updated version an AI model which can be used to generate high-resolution wind, rainfall, and sea level pressure predictions over long simulations.
Several updates to the model training and prediction processes are shown to improve its ability to simulate the most severe events.
We show how HiRO-ACEv2 could be used to study the prevalence and intensities of extreme events in high-resolution simulations, at a fraction of the computational cost of running an equivalent numerical model.

%
%

\section{Background}

Physics-based global storm-resolving models (GSRMs) simulate the atmosphere at kilometer scales ($\lesssim5$ km horizontal resolution), providing sufficient spatial resolution to resolve the severity and locality of extreme events and climate impacts \citep{stevens2019,wille2025}.
They can also directly resolve key atmospheric processes such as deep convection and interactions with complex topography that are only approximated in coarse-resolution ($\mathcal{O}$(100) km scales) global climate models (GCMs), which makes their projections more accurate than those of GCMs.
However, the high computational cost associated with running a single GSRM for multiple years precludes their use for aggregating climate statistics or studying the prevalence of extreme events over long timescales and large ensembles. 

One way to obtain kilometer-scale predictions without running a GSRM is to downscale coarse-resolution GCM data \citep{giorgi2019,lucaspicher2021,rampal2024}. The term ‘downscaling’ may refer to multiple methods by which one may generate predictions at the kilometer-scale using coarser-grained inputs. 
Dynamical downscaling utilizes a lower resolution GCM state as boundary conditions for running a regional, high-resolution model; the reduction in computational resources comes from the much decreased area of the globe that is simulated at high resolution.
Statistical downscaling makes use of learned relationships between coarse-resolution fields and signals at the local level. 
This can be more computationally efficient than dynamical downscaling, but prediction quality suffers if the assumption of stationarity no longer holds.
In recent years, the range of statistical downscaling techniques has massively broadened from classical statistical fitting methods to include artificial intelligence (AI) approaches \citep{rampal2024,chun2026}.

A substantial body of work covers the use of generative modeling for downscaling purposes \citep{najafi2026}. 
In particular, diffusion-based generative modeling has proven to be a powerful tool for downscaling coarse-resolution fields to horizontal resolutions $\lesssim10$ km \citep{saoulis2025,mardani2024,brenowitz2025,wan2026}. 
Diffusion-based approaches allow researchers to stochastically generate ensemble predictions to provide uncertainty estimates for high-resolution outputs, while maintaining realistic small-scale spatial structure and sharpness \citep{lopezgomez2025,brazidec2026}.

Several works share similar strategies to \hiroace in downscaling coarse-resolution states to produce high-resolution climate data suitable for studying extreme events.
CBottle \citep{brenowitz2025} probabilistically generates coarse-resolution states from $\sim$100~km ERA5 reanalysis or coarsened ICON GSRM data, conditioned on the target dataset, datetime, and sea surface temperature, then applies a second model to superresolve to a final $\sim$5 km state consistent with the original ICON data.
The GenFocal framework \citep{wan2026} debiases and downscales 1.5\degree~($\sim$150 km) resolution Community Earth System Model states to 0.25\degree($\sim$25 km) resolution states trained to be consistent with ERA5 reanalysis. It reproduces its target's cyclogenesis and intensity statistics as well as the heat extremes.
\citet{guan2026} run a $\sim$300 km autoregressive emulator trained on ERA5 and downscale resulting outputs to $\sim$28 km resolution, providing another example of how lightweight coarse-resolution emulators can be used to generate high-resolution climate projections.
Here, we introduce \hiroace, which autoregressively simulates climate at 100~km resolution before downscaling to 3~km resolution. The unique features of \hiroace are 1) the global autoregressive model is extremely fast and can simulate atmospheric variability on a wide range of time scales, and 2) the downscaling model operates at a 32$\times$ super-resolution factor, allowing reproduction of X-SHiELD data at its native resolution.

We note that the term `downscaling' is often be used to describe a process that includes a bias correction step for the coarse inputs before super-resolving to the final resolution.
Throughout this work, we use the term to refer to both the prediction of high-resolution outputs from coarsened versions of the target 3~km state (the `perfect prediction' case) as well as the generation of high-resolution fields using inputs from a free-running GCM emulator. We will show that even though our downscaling model is only trained in the perfect prediction setting, when used in the latter setting our autoregressive emulator's outputs are sufficiently close to the target climate that we do not require a bias correction step.

\subsection{\hiroace}
The \hiroace framework described in this work was originally introduced in \perkins. 
It consists of two distinct, separately trained AI models: a stochastic version of the Ai2 Climate Emulator (\ace) and a generative diffusion-based downscaling model (High Resolution Output, or HiRO).
Both components are trained using coarse- and high-resolution versions of the same 10 year X-SHiELD GSRM dataset, so that the coarse emulator outputs are consistent with the input distribution that the downscaling model is trained on.
\hiroace runs in two stages: \ace autoregressively predicts 100~km coarse-resolution states at 6 hour frequency, which are then provided as inputs to HiROv2 to downscale to 3~km resolution.

\perkins{} performed a comprehensive analysis of HiRO-ACE's fine-resolution precipitation statistics and biases.
In this work, we extend the framework beyond its original precipitation-only outputs and generate multivariate downscaled predictions for precipitation, 10~m eastward and northward winds, and sea level pressure, enabling the study of a wider range of phenomena.
We emphasize the aspects of our training methodology that improve HiRO's ability to represent the far tails of the target distributions.
In order for this framework to be useful for studying the prevalence and distribution of extreme events, two conditions must be met: the statistics of the \ace emulator's autoregressive predictions should match those of the coarsened reference X-SHiELD data, and the HiROv2 downscaling model's outputs should do the same with respect to the 3~km X-SHiELD data. 
Over the following sections, we will show that these two conditions are met for our set of multivariate outputs. 
Section \ref{sec:data} describes the datasets used for training the our \ace and HiROv2 models.
Section \ref{sec:hiro_perfect_pred} details the training methodology for the downscaling HiROv2 component with particular attention paid to the updates over v1 which improve its long-tail predictions, and discusses its ability to reproduce accurate global statistics over time, tropical cyclones, and single timestep events.
Section \ref{sec:ace_coarse_emulator} describes \ace emulator training and shows that the statistics of 10-year aggregated \ace outputs and tropical cyclogenesis are similar to the 100~km X-SHiELD reference.
Finally, Section \ref{sec:hiroace_results} demonstrates how the \hiroace pipeline can generate high-resolution climatology, long-tailed statistics, and realistic event snapshots over a decade-long free-running inference period.

\section{Data}
\label{sec:data}
\subsection{X-SHiELD 3~km and 100~km reference dataset}
This work uses output from the same 11-year GFDL X-SHiELD (eXperimental System for High-resolution prediction on Earth-to-Local Domains) simulation initialized on 2013-01-01 used in \perkins. We refer the reader there for a detailed description of the simulation configuration, computational cost, and data processing steps, but summarize the main points here. 

Internally X-SHiELD uses a $\sim$3.25~km resolution (C3072) cubed sphere grid \citep{cheng2022,harris2023}. For conciseness, we hereafter refer to this as `3~km' resolution.
Through a combination of online and offline remapping steps, data on the 3~km grid is coarsened by roughly a factor of 32 to a 1\degree ($\sim100$~km) Gaussian grid; this coarse-resolution dataset is used in both the downscaling (HiRO) and GCM emulator (ACE2S) training. 
For downscaling, a selection of fields are also output on the native $\sim$3.25~km grid and are conservatively remapped offline to a custom 5760$\times$11520 fine-resolution grid, constructed by uniformly refining the 180$\times$360 Gaussian grid by a factor of 32 along the latitude and longitude dimensions. The resulting paired coarse and fine-resolution datasets are used to train HiRO. The coarsened data is used in the fine-tuning step of ACE2S training.

The train-test split follows that in \perkins.  The first year of the simulation is excluded as a spin-up period and both the downscaling model and the coarse-resolution \ace emulator use the subsequent 9 years of the dataset (2014-2022) for training, leaving the last year (2023) as the holdout dataset.

It should be noted that the surface precipitation rate is averaged over the 6 hour interval preceding the snapshot timestamp, while the 10~m winds and sea level pressure fields are instantaneous snapshots at that timestamp. 
Thus, the precipitation field in this work has a more lagged covariance with winds and pressure than if it were an instantaneous precipitation rate.

\subsection{SHiELD 100~km pretraining dataset}
\label{sec:ace2s-shield-plus-dataset}

The ACE2S emulator used in this work differs from \perkins{} in its pretraining dataset.  Rather than start from a checkpoint trained on ERA5 reanalysis \citep{Her2020}, this work starts from a checkpoint trained on output from a 100 km resolution (C96) version of SHiELD.  X-SHiELD and SHiELD share the same base simulation code, but are run using different configurations for their respective resolutions \citep{harris2020, cheng2022}.  While their differences in resolution and configuration of physical parameterizations naturally lead to different results, the two models are more similar to each other than they are to ERA5 with respect to some important aspects of macroscale climate, like their distribution of tropical cyclones, making SHiELD a more compatible source of pretraining data.

Our base checkpoint is the ACE2S-SHiELD+ model described in \citet{clark2026}.
This model was trained on 90 years of data derived from a collection of historical climate, idealized climate change, and system identification simulations designed to disentangle the independent effects of SST and CO$_2$. We refer the reader to \citet{clark2026} for the full details of this dataset, but for the purposes of this study, what is most relevant is that this represents a base of 10x as many training samples as that available from the X-SHiELD dataset alone. Prior to fine-tuning on the X-SHiELD dataset, we adapt this checkpoint using this SHiELD dataset to include sea level pressure as an output.

\section{HiRO: Multivariate downscaling of winds, precipitation, and sea level pressure}
\label{sec:hiro_perfect_pred}

\subsection{Model training}
\label{subsec:model_training}
The multivariate HiROv2 model used in this work uses the same architecture as the precipitation-only downscaling model released in \perkins, but with added input and output channels for the additional fields. 
Its inputs are the coarse-resolution surface precipitation rate, 10~m eastward and northward winds, sea level pressure, and fine-resolution static surface height and land fraction. 
The model simultaneously outputs precipitation, 10~m winds, and sea level pressure fields downscaled to 3~km resolution.  

The downscaling model uses the CorrDiff residual correction framework \citep{mardani2024} to stochastically generate downscaled fields. However, as in \citet{lopezgomez2025}, the base prediction is a simple bicubic interpolation of the input coarse-grid fields to the fine-resolution grid rather than a trained regression UNet as in \citet{mardani2024}. 
A diffusion model then predicts high-resolution residual output fields, which when added to the deterministic base prediction produce the final ensemble of downscaled outputs. 
Predictions are generated over 16\degree$\times$16\degree patches; for inference over larger areas we generate and composite multiple independent patch predictions (for more details, see Appendix \ref{app:hiro_inference}).

A high-level overview of the diffusion modeling procedure is given here as context for our training updates; for a full description of the specific Elucidated Diffusion Models (EDM) framework used here, see \citet{karras2022}.
We train HiRO to learn the residual corrections $\mathbf{y}$, the difference between the normalized 3~km full-field target and interpolated normalized coarse channels.
Inputs $\textbf{x}$ are the concatenation of the interpolated coarse input fields and fine-resolution static input fields.
A noise level $\sigma$ is drawn from a distribution $p(\sigma)$ and Gaussian noise $\mathbf{n}\sim\mathcal{N}(0, \sigma^2 \mathbf{I})$ is added to $\mathbf{y}$.
A denoiser $D_\theta$ is trained to output the unnoised $\mathbf{y}$ when conditioned on noise level $\sigma$ and $\mathbf{x}$. The loss being minimized is
\begin{equation}
  \mathcal{L} = \mathbb{E}_{\sigma,\,\mathbf{y},\,\mathbf{n}}
  \left[\lambda(\sigma)\,\frac{1}{N}\sum_{c,i} w_c
\left(D_\theta(\mathbf{y}+\mathbf{n};\sigma,\mathbf{x})_{c,i}-y_{c,i}\right)^2\right],
  \label{eq:diffusion_loss}
\end{equation}
where $c$ and $i$ indices cover output channels and grid points respectively, $N$ is the total number of grid points over channels, $w_c$ is a per-channel weight, and $\lambda(\sigma)$ is a noise-dependent weight (further described in Section \ref{subsubsec:moe_denoisers}). At inference time, each downscaled sample is generated by starting with pure noise at the scale $\sigma_{max}$, and iteratively denoising over decreasing noise levels down to $\sigma_{min}$ over a number of steps.

We made several adjustments to the HiROv1 \citep{perkins2026} noise and data sampling, and modified the model inference to partition denoising steps across two separately-trained denoisers.
These updates, described below, significantly improved the ability of the diffusion model to generate outputs in the tails of the target distribution and thereby improved the model's skill in extreme predictions.
For full details of the model training and inference procedures, see \ref{app:hiro_details}. Differences in training procedure between HiRO versions 1 and 2 are summarized in Table \ref{tab:hiro-v1-v2}.

\subsubsection{Log uniform noise sampling improves extreme precipitation}
\label{subsubsec:log_uni_noise}
The precipitation-only HiROv1 downscaling model used training methods and parameter settings that were very close to those described in the original EDM framework \citep{karras2022}. The image generation task described in \citet{karras2022} occurs in red-green-blue space, where the normalized data is well-bounded. In contrast, precipitation is a heavy-tailed variable. The most extreme precipitation samples in the dataset---which are of the most interest to us in this work---have z-scores $\gg$10. During EDM training, the noise sampling distribution $p(\sigma)$ is a log normal distribution. If the high-noise tail of $p(\sigma)$ does not sufficiently cover the highest z-score data values during training, then the denoiser will be unaccustomed to encountering pure noise at the start of the denoising process in combination with the coarse inputs associated with the most extreme events, which should be avoided.

The values of $p_{mean}=-1.2,~p_{std}=1.8$ used in \perkins~ to parameterize the log normal $p(\sigma)$ were chosen to extend the width of the distribution to account for high precipitation values.
Upon further iteration, we found that using a log uniform noise distribution for $p(\sigma)$ as described in \citet{brenowitz2025} during training improved the generation of extreme precipitation (Fig. \ref{fig:methods_ablation_histograms}). 
The choice of training noise distribution had a significant effect on precipitation, but not on the wind or pressure extremes. This is not surprising, as precipitation is the output field with the highest dynamic range by far.

Noise was sampled during training from a distribution with uniform probability in log space between ($\sigma_{min}, \sigma_{max}$). As described in more detail below, two separate denoisers are trained using noise ranges ($\sigma_{min}, \sigma_{max}$)=\{(0.005, 200), (200, 2000)\}.
More details on how the distribution bounds were chosen are given in Appendix \ref{app:hiro_training}.

\subsubsection{Separate denoisers for high and low noise schedule ranges}
\label{subsubsec:moe_denoisers}
Adding more output channels to the original precipitation-only diffusion model introduces more complexity to the original single-output denoising task. As mentioned above, precipitation requires a higher starting noise due to its long-tailed nature. 
A single denoiser trained to handle the full noise range at inference was unable to generate the full magnitude of winds in the most extreme events. A blunt, yet effective, method of improving the model's ability to capture extremes was to train two separate denoisers to handle the high ($200<\sigma\leq2000$) and low ($0.005\leq\sigma\leq200$) noise ranges during prediction.
This is similar to the approach taken in \citet{brenowitz2025} though our motivation differs somewhat from theirs (which was to avoid overfitting at high noise levels).
The threshold value of $\sigma=200$ used to route the inputs to the appropriate denoiser during inference was chosen based on the noise value that would have been an appropriate $\sigma_{max}$ value if only winds and pressure were being downscaled. The following adjustments were made to the low range denoiser only.

The loss contribution $w_c$ from the precipitation channel was multiplied by a factor of 0.2. This loss weighting prevents precipitation from dominating the loss when extreme samples are scored during training, as the z-scores for precipitation in strong storm events can be an order of magnitude larger than the most extreme wind and pressure z-scores.

In Equation \ref{eq:diffusion_loss}, sample loss is weighted by a factor $\lambda(\sigma)$ in order to balance the variance in target magnitudes over the range of preconditioning noise.
\begin{equation}
    \lambda(\sigma) = \left( \frac{\sigma^2 + \sigma_{data}^2}{(\sigma\cdot\sigma_{data})^2} \right)^a
\end{equation}
In the EDM training procedure described in \citet{karras2022}, $a=1$. 
The constant $\sigma_{data}$ is usually taken as the channel standard deviation; here we use the same value $\sigma_{data}=1.0$ (standard normalization is applied across all channels).
We found that when using the standard loss weighting with $a=1$, weighted loss for the lowest noise samples ($\sigma\lesssim 0.05$) in the low range denoiser was approximately an order of magnitude larger than for samples with $\sigma\gtrsim1$. 
This led to an overemphasis on learning to denoise the smallest-scale features, at the expense of model skill on larger-scale features and extreme values.
To balance the loss contributions across the training noise range we used a value of $a=0.75$ when training the low range denoiser. The high range denoiser is trained using the standard EDM loss weighting with $a=1$.

\begin{figure}[h]
    \centering
    \includegraphics[width=1.0\linewidth]{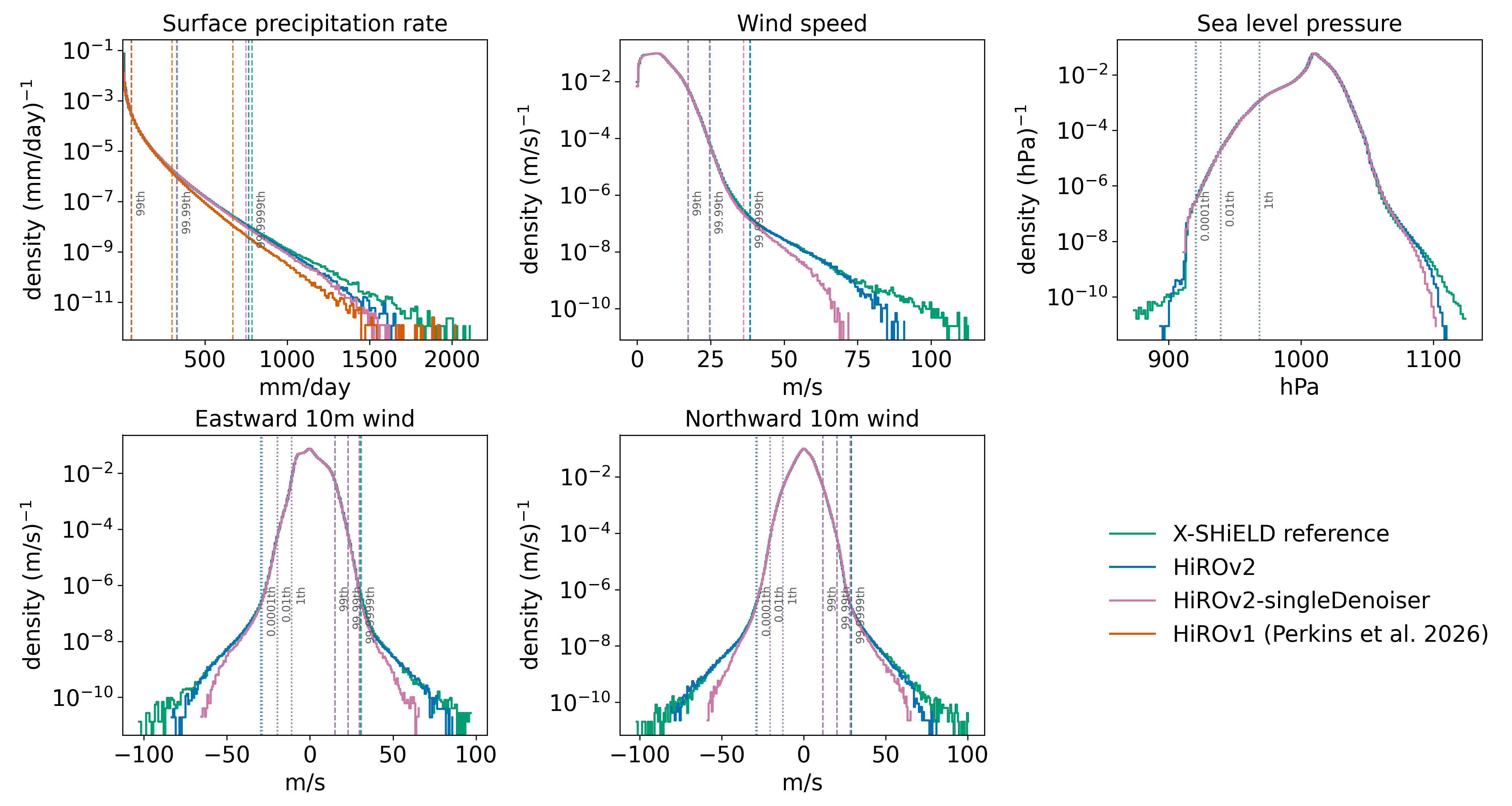}
    \caption{
        Aggregated histograms of surface precipitation, 10~m winds, and sea level pressure fields between 65\degree N--65\degree S over 2023 demonstrating how adaptations to diffusion modeling led to improved downscaling of multivariate tail statistics.
        `HiROv1' refers to downscaled output using the precipitation-only model released in \perkins, which was trained using a log normal noise distribution.
        `HiROv2-singleDenoiser' uses a single denoising diffusion model for all multivariate outputs trained using a log uniform noise distribution, and does not include any of the training updates described in \ref{subsubsec:moe_denoisers}.
        `HiROv2' is the final model used to generate the main results in this work, and includes the full set of updates described in \ref{subsubsec:moe_denoisers} and \ref{subsubsec:log_uni_noise}.
        Dashed lines indicate \{99, 99.99, 99.9999\}th percentile values for each distribution.
    }
    \label{fig:methods_ablation_histograms}
\end{figure}

Separating the model into low and high range denoisers and using the aforementioned loss weightings in the low range denoiser substantially improved the model's ability to generate the most extreme wind and sea level pressure values in the target distribution. Fig. \ref{fig:methods_ablation_histograms} shows the histograms and extreme percentile values for the held-out 2023 year from several iterations of the HiROv2 downscaling model and the reference 3~km X-SHiELD. Note that wind speed is not a direct model output, but is derived from the eastward and northward 10~m wind outputs.
Comparison of the precipitation histograms from `HiROv1' to `HiROv2-singleDenoiser' demonstrates the effect of changing the training noise sampling distribution.\footnote{
    This is not a true ablation as the `HiROv2-singleDenoiser' model also includes oversampling of tropic patches and additional multivariate outputs. These differences do not substantially affect the global precipitation histogram (not shown).
}
The effect of training separate denoisers and adjusting the low range noise weighting are most evident in the wind and sea level pressure distribution tails when comparing `HiROv2-singleDenoiser' to the final model used in the rest of this work, `HiROv2' (though they also yield a small improvement in the 99.9999th percentile for precipitation). Overall, the final model is able to generate downscaled predictions that closely match the reference X-SHiELD distribution out to the 99.9999th percentile of all output fields.

\subsubsection{Oversampling tropical latitudes}
Since tropical cyclones are rare events in the training dataset, we weighted the sampling of training data to oversample patches with center latitudes within $\pm30$\degree by a factor of 8 relative to the rest of the domain. 
We found that the model skill in predicting extreme tail values of precipitation and winds during tropical cyclone events was improved by this sampling approach, with no significant degradation to the performance on time mean climatology or extratropical events.

\subsection{Downscaling tropical cyclones: aggregated statistics and morphology}
\label{subsec:perfect_pred_tc_stats}

Fig. \ref{fig:methods_ablation_histograms} demonstrates that the multivariate HiROv2 downscaling model is able to reproduce aggregated target distributions. Here, we present results from downscaling individual events and evaluate the model's ability to generate realistic tropical cyclones.

The set of tropical cyclone events analyzed here was selected from the held-out 2023 year of 100~km X-SHiELD dataset by running the TempestExtremes detection software \citep{ullrich2021} using the same methodology as described in \citet{watt-meyer2025}.
The timestep with the maximum coarse-field 10~m wind speed in each cyclone track was selected and downscaled to an 8-member ensemble.
In total, 72 tropical cyclones were detected and downscaled from the 2023 coarsened X-SHiELD data. 

We compute the integrated kinetic energy (IKE) for each cyclone snapshot as a measure of the storm's destructive potential \citep{powell2007}. 
We use IKE as our intensity metric for high-resolution comparisons against the reference because it takes into account more grid cells than the single-point measure of maximum wind.
IKE is calculated as
\begin{equation}
    IKE=\int_A \frac{1}{2}\rho U^2 ~\mathbf{1}_{\{U \ge 18\text{m/s}\}}\ dA.
\end{equation}
The integrated area $A$ is a 1 meter thick, 8\degree$\times$8\degree~box horizontally centered on the storm pressure minimum and vertically centered at 10 meters, discretized into the approximately 3~km horizontal grid.
The density of air within $V$ is taken as a constant 1 kg/m$^3$.
$U$ is the 10 m wind speed field; the integral is over regions with $U>18$ m/s (tropical storm force winds). 

\begin{figure}[h]
    \centering
    \includegraphics[width=0.75\linewidth]{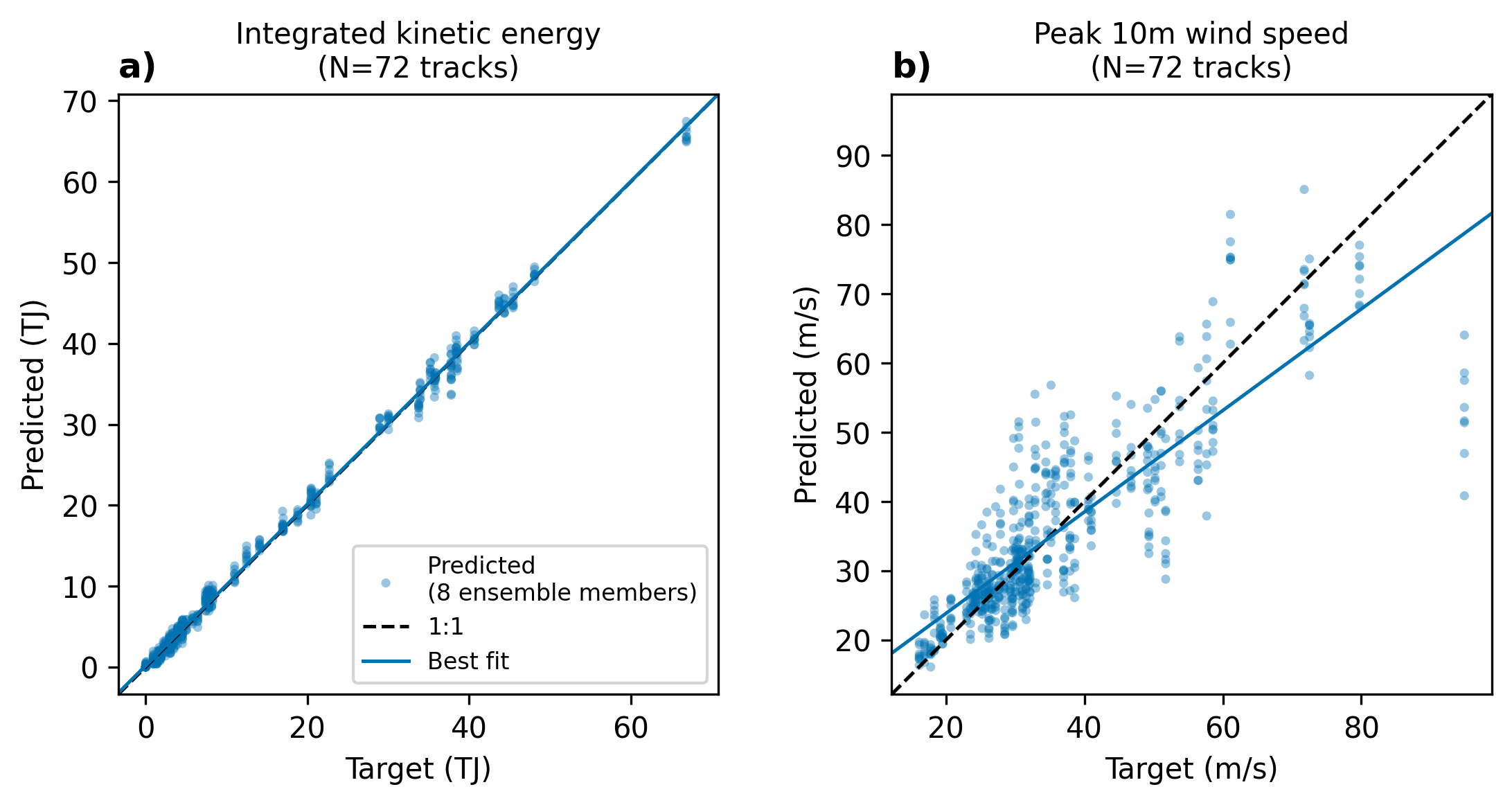}
    \caption{a) HiRO-predicted vs. target values of integrated kinetic energy for the set of tropical cyclones from 2023. Points that sit along a vertical line at the same target value correspond to generated samples for the same downscaled snapshot. b) Same as (a), but for maximum 10~m wind speed.}
    \label{fig:perfect_pred_ike_scatter}
\end{figure}

Fig. \ref{fig:perfect_pred_ike_scatter}a shows the predicted versus target IKE over the set of tropical cyclones. The storm intensities predicted by HiROv2 downscaling track the target intensities very well, with a RMSE of 0.7 TJ over the set. Even for the strongest cyclones with maximum wind speeds $>$50 m/s, the mean IKE RMSE is $<$1 TJ. In contrast, the spread in predicted maximum wind speed about the target 1-to-1 line is much larger (Fig. \ref{fig:perfect_pred_ike_scatter}b)-- predicting a single extreme high-resolution value is a much more difficult task than generating an extended system with the correct spatially-integrated IKE. 
The systems with the strongest winds in the target dataset ($\gtrsim40$ m/s) tend to have their maximum wind speeds under-predicted by HiROv2.

\begin{figure}[H]
    \centering
    \includegraphics[width=1.0\linewidth]{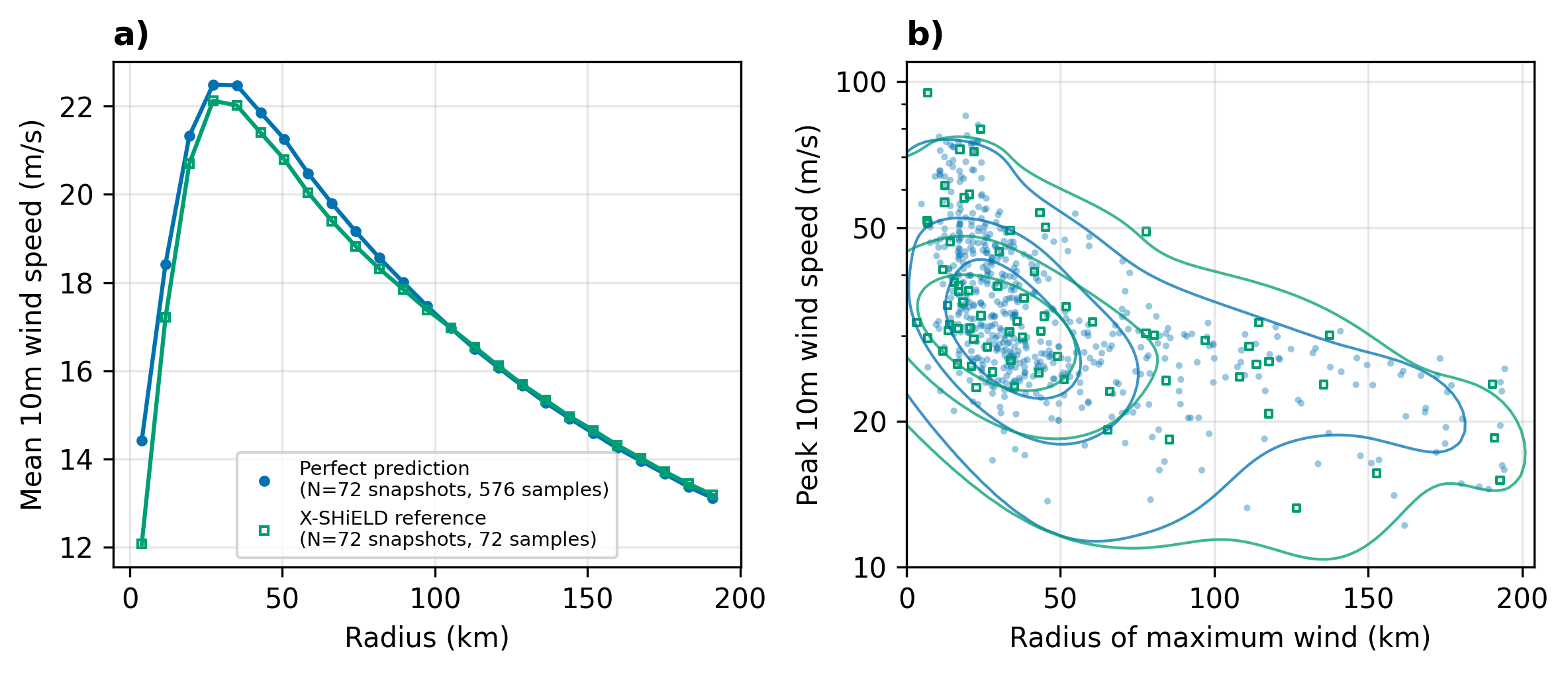}
    \caption{
        a) Mean radial wind profile of 3~km resolution tropical cyclones in the 2023 year for HiROv2 perfect prediction downscaled samples that used 100~km coarsened X-SHiELD fields as inputs (blue), and of the corresponding target 3~km X-SHiELD target (green). Bins are centered on the minimum pressure grid cell.
        b) Scatter plot and density contours showing the relation between cyclone maximum wind speed and radius of maximum wind across the set of tropical cyclones.    
    }
    \label{fig:perfect_pred_radial_wind_profile}
\end{figure}

Mean radial wind profiles are computed by azimuthally averaging 10~m wind speed in radial bins centered on each snapshot's pressure minimum.
The reference set and downscaled predictions have very similar shape, peak amplitude, and radius of maximum wind (Fig. \ref{fig:perfect_pred_radial_wind_profile}a). 
The largest difference in the profile occurs in the innermost radial bin, where the perfect prediction mean wind speed is $\sim$20\% higher than the reference. 
This is largely attributable to the fact that the HiRO-downscaled tropical cyclones are not as radially symmetric as their targets (a qualitative example of this is visible in Fig. \ref{fig:perfect_pred_Phl_tc}c), which results in higher mean wind speed values closer to the center when averaging azimuthally within those radial bins. 
The full two-dimensional distributions of the reference and predicted maximum wind versus radius of maximum wind demonstrate that the downscaled storms follow the same size and wind speed relationship as the reference across the range of data tested (Fig. \ref{fig:perfect_pred_radial_wind_profile}b).

\subsection{Case study: downscaling a strong tropical cyclone}
\label{subsec:perfect_pred_tc_example}

\begin{figure}[h]
    \centering
    \includegraphics[width=1\linewidth]{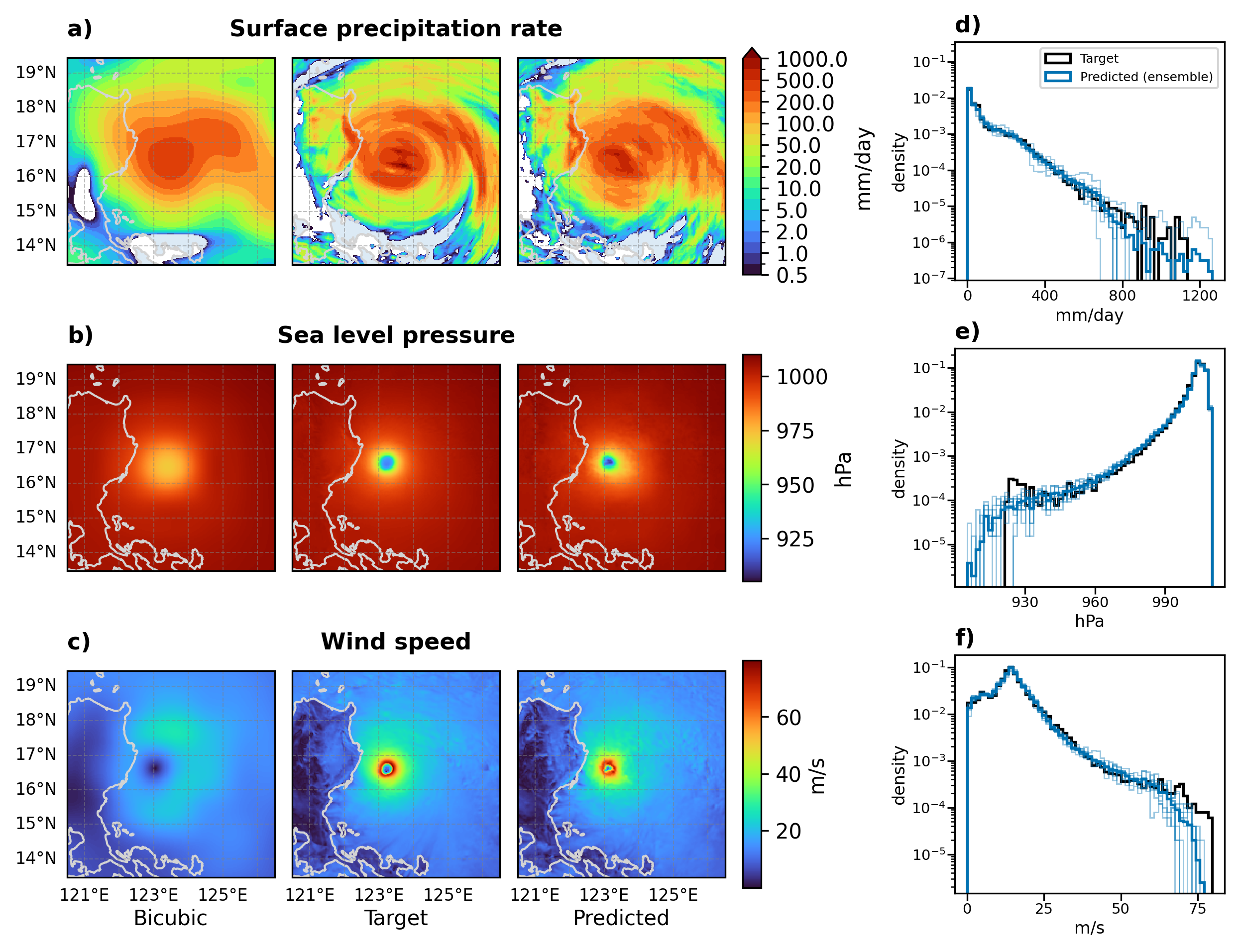}
    \caption{
        Bicubic-interpolated 100~km resolution, target 3~km X-SHiELD, and predicted 3~km HiROv2 downscaled snapshots of a tropical cyclone near the Philippines on 2023-10-27. Histograms are shown in the right column are for the target and predicted distributions of each output variable within the plotted region.
        Bold blue lines in the histograms denote the ensemble mean prediction, with fainter blue lines showing the pdf for each of the 8 generated ensemble members.
    }
    \label{fig:perfect_pred_Phl_tc}
\end{figure}

Fig. \ref{fig:perfect_pred_Phl_tc} a-c show examples of downscaled single-timestep precipitation, pressure, and wind fields in a strong tropical cyclone in X-SHiELD with reference wind speeds exceeding 75 m/s (270 km/h). For visual clarity, only the first sample out of 8 generated samples is shown in the downscaled prediction column, but the full 8 member prediction is provided in Fig. \ref{fig_sup:perfect_pred_Phl_tc_ensemble}. The results described here in the example sample generally hold true across the rest of the ensemble.

HiROv2 generates the central precipitation maximum, outer bands, and increased rainfall over land, though the banded structure in the outer precipitation field is somewhat less coherent in the prediction than in the reference. 
While the envelope of predicted maximum precipitation in the histogram encompasses the reference target, there is a large spread in the ensemble prediction; we also observe comparably large dispersion in the downscaled precipitation histogram tails for the other tropical cyclones from the holdout set. 
HiROv2 resolves the general annular shape of the sea level pressure and 10~m wind speed fields, as well as the drop in wind speed over land. However, these fields are less radially symmetric within the inner $\sim1$\degree than the reference storm. 
Additionally, the histogram of sea level pressures shows that HiROv2 struggles with reproducing the steep cutoff at the minimum pressure within the storm's eye.
Maximum wind speed is biased low across the ensemble in this particular example, consistent with our previously stated finding that HiROv2 develops a low bias in downscaled maximum wind speed in the reference storms with the strongest winds (Fig.~\ref{fig:perfect_pred_ike_scatter}b).

Since consecutive input snapshots are downscaled independently, a time series of high-resolution predictions along a tropical cyclone track has less temporal coherence than the corresponding X-SHiELD reference track. Taking an ensemble mean helps to smooth the variance between timesteps, but a lack of temporal coherence is still evident in individual ensemble members (Appendix Fig. \ref{fig_sup:perfect_pred_tc_time_series}).

\subsection{Case study: Downscaled winds over topography}
\begin{figure}[h]
    \centering
    \includegraphics[width=1.0\linewidth]{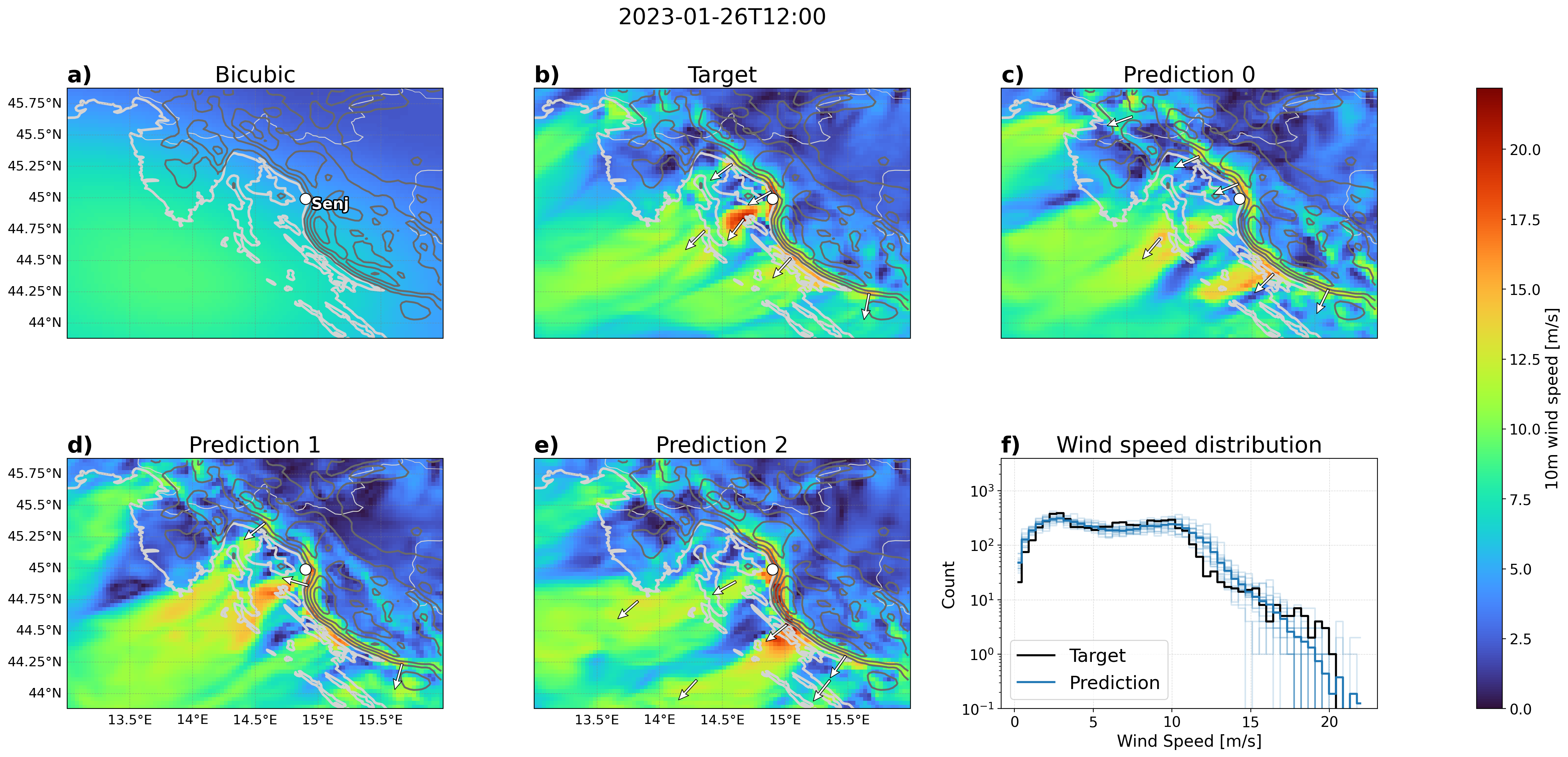}
    \caption{
    (a) Bicubic interpolation of 100~km X-SHiELD input, (b) 3~km X-SHiELD target, (c-e) generated samples of 3~km HiROv2-downscaled 10~m wind speed fields for a snapshot of a Bora wind event in Croatia.
    Topography contours are indicated as grey lines over land. Arrows at the locations of strongest winds indicate the wind vector at the cell they originate on. 
    (f) Histogram of target and predicted 3~km wind speeds within the plotted region. The bold blue line denotes the ensemble mean prediction, with fainter blue lines showing the 16 individual ensemble members.}
    \label{fig:perfect_pred_bora}
\end{figure}

In addition to tropical cyclones, another practical use case for high-resolution downscaled climate data is to study the intensity and prevalence of strong wind events.
As an example, we show snapshots of the reference and HiRO-predicted 10~m wind speed fields from the held out 2023 data of a Bora event over Croatia (Fig. \ref{fig:perfect_pred_bora}) where cold katabatic winds flow downslope into the Adriatic Sea.
For visual conciseness, only the first three generated sample maps are shown. 
While the locations of the highest winds vary across the generated ensemble, the HiRO-generated wind vectors over land exhibit physically realistic downslope flows.
The 10~m wind speed distribution over all generated samples agrees well with the reference (Fig. \ref{fig:perfect_pred_bora}f).

\section{ACE2S-X-SHiELD: a stochastic emulator of coarsened 100~km X-SHiELD}
\label{sec:ace_coarse_emulator}

\subsection{Model training}
The coarse \ace emulator used in the first step of \hiroace is trained following a similar methodology as described in \perkins: an \ace checkpoint that was pretrained on a different, larger dataset is fine-tuned on nine years of coarsened X-SHiELD data. The fine-tuning step results in an emulator that can generate 100~km output distributions that are consistent with the coarsened X-SHiELD data used in both the emulator tuning and the downscaling training. Full details of the training setup, variable list, and parameters are given in \ref{app:ace-training}. 

As noted in Section~\ref{sec:ace2s-shield-plus-dataset}, a difference in this work from \perkins{} is that the model is pretrained on output from a collection of coarse-resolution SHiELD simulations rather than ERA5 reanalysis. We tested the sensitivity to this choice and found that while both pretraining datasets resulted in similar precipitation biases in the final fine-tuned model, the base checkpoint trained on output from SHiELD resulted in a tropical cyclone climatology more similar to that of the target X-SHiELD dataset.
Hereafter, usage of the term `\ace' refers to the final emulator used in this work, which was an ACE2S-SHiELD+ model from training done as part of \citet{clark2026}, fine-tuned to additionally predict sea level pressure, and further fine-tuned on the X-SHiELD dataset.

\subsection{\ace climatology}

\begin{figure}[h]
    \centering
    \includegraphics[width=1.0\linewidth]{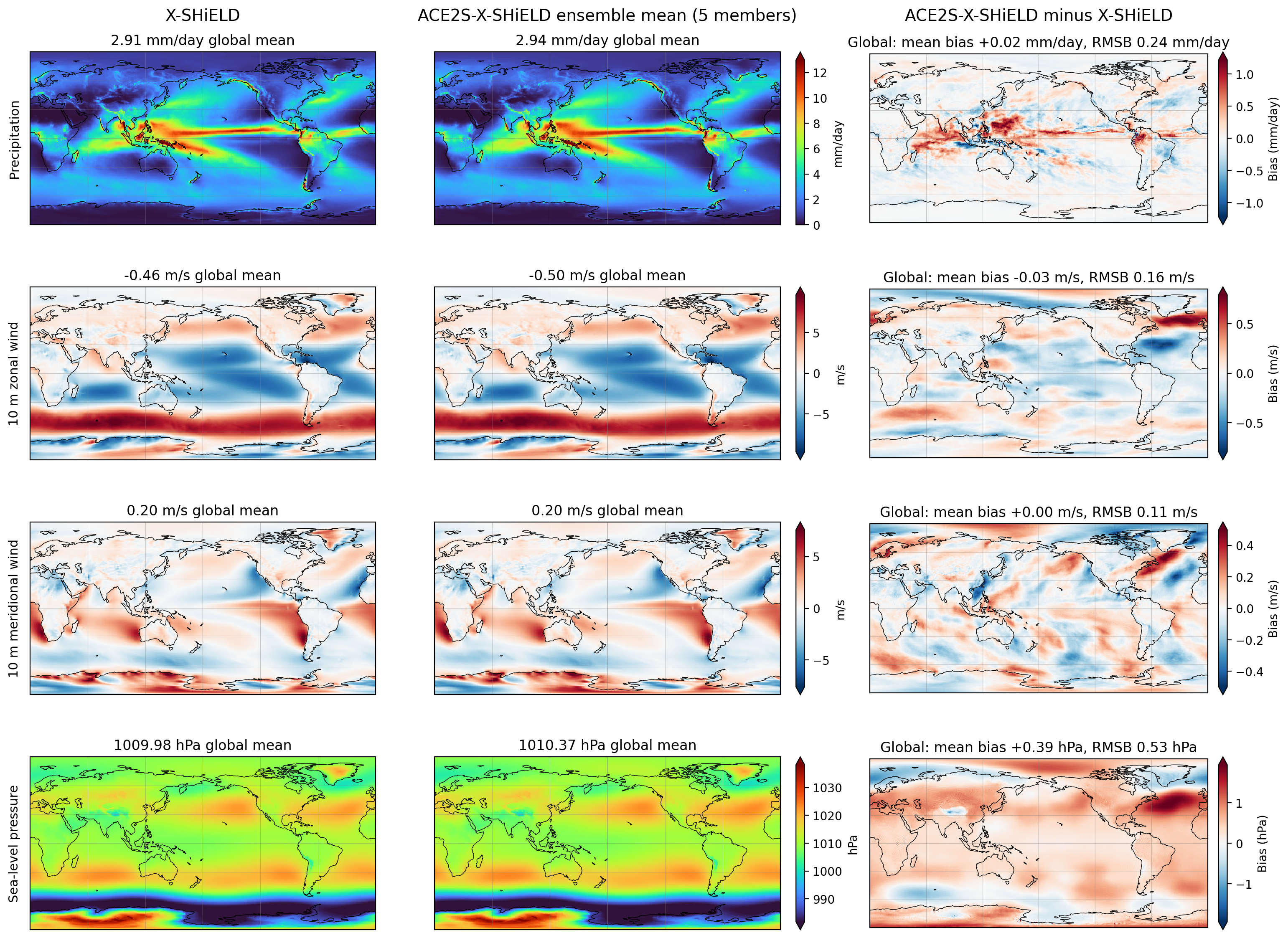}
    \caption{
        10 year mean climatologies for precipitation, 10~m zonal and meridional winds, and sea level pressure for 100~km X-SHiELD (left) and \ace (middle). The time mean bias pattern of \ace with respect to X-SHiELD is shown in the right column.
    }
    \label{fig:ace2s_10y_climatology}
\end{figure}

We perform five free-running \ace 10-year runs from 2014--2023 from an ensemble of initial conditions, forced with the same sea surface temperatures, sea ice, and CO$_2$ concentration from the corresponding years of the reference X-SHiELD simulation.
Fig. \ref{fig:ace2s_10y_climatology} shows the time- and ensemble-mean climatology of the resulting dataset.
The root mean squared biases (RMSBs) of the time-mean precipitation and zonal winds (sea level pressure was not an output in previous models) are similar to or smaller than in previous deterministic versions of ACE trained on ERA5 or SHiELD data \citep{watt-meyer2025}. Compared to the version of ACE2S-X-SHiELD used in \perkins (Table \ref{tab:ace2s_rmsbs}), the mean RMSB over ensemble members is comparable to but slightly higher in this work.

\begin{table}[h]
  \centering
  \begin{tabular}{lcc}
    \toprule
    & \multicolumn{2}{c}{Root mean squared bias} \\
    \cmidrule(lr){2-3}
    Field & \perkins & This work \\
    \midrule
    Precipitation (mm/day)     & $0.23$  & 0.28 ($0.27$-$0.30$) \\
    10~m zonal wind (m/s)      & $0.18$  & 0.20  ($0.18$-$0.21$)\\
    10~m meridional wind (m/s)& $0.13$ & 0.14  ($0.13$-$0.15$)  \\
    Sea level pressure (hPa)   & not predicted  & 0.59  ($0.55$-$0.62$) \\
    \bottomrule
  \end{tabular}
  \caption{Global root-mean-square bias (RMSB) of the 2014--2023 time-mean 100~km resolution 
    fields relative to X-SHiELD. 
    Values from this work are the mean of the five ensemble members' individual RMSBs, with the range
    across members given in parentheses. Values from \perkins{} are from a single
    ensemble member.}
 \label{tab:ace2s_rmsbs}
\end{table}

The precipitation bias pattern in this version of \ace-- notably in the tropical west Pacific and South America-- is qualitatively similar to the time-mean bias in spatial distribution and magnitude to the version used in \perkins.
As these two versions of \ace are pretrained on different datasets (ERA5 in \perkins and C96 SHiELD here) with very different precipitation biases relative to 100~km X-SHiELD, these broad similarities suggest that most of the final models' precipitation biases develop during the fine-tuning step on X-SHiELD data.

\subsection{Tropical cyclone climatology in coarse-resolution ACE}
\label{subsec:ace_tc_climatology}

\begin{figure}[h]
    \centering
    \includegraphics[width=1.0\linewidth]{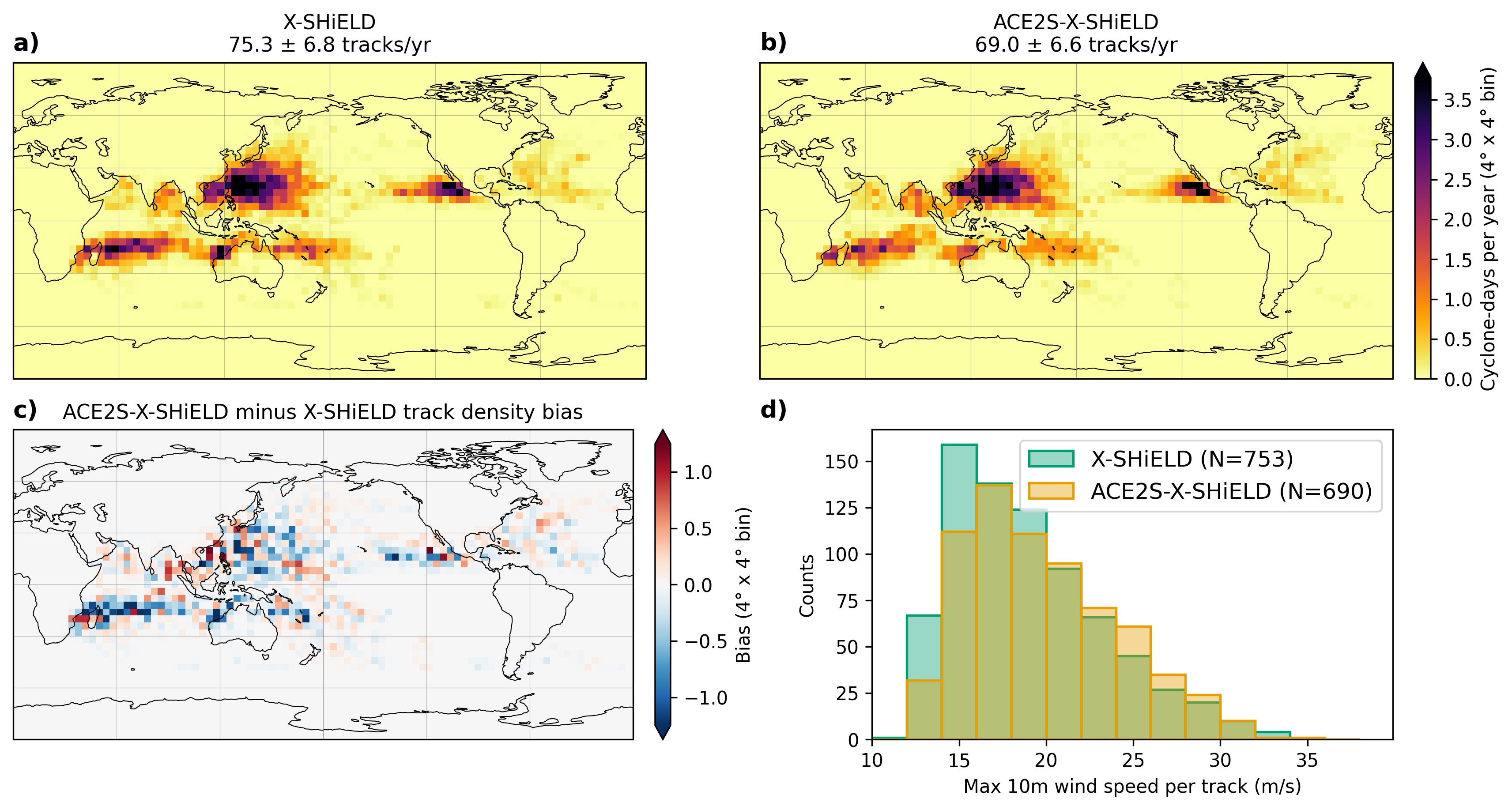}
    \caption{(a, b) Tropical cyclone track densities (per 4$^\circ\times$4\degree bin) in 10 years of 100~km-coarsened X-SHiELD reference and free-running \ace simulations, respectively. (c) \ace track density bias. (d) Histogram of maximum coarse-resolution cell 10~m wind speeds for all tropical cyclones detected in 10 years of X-SHiELD and \ace simulations. }
    \label{fig:coarse_tc_tracks}
\end{figure}

Tropical cyclone tracks are counted in the reference and 100~km \ace output using the TempestExtremes selection criteria as described in Section \ref{subsec:perfect_pred_tc_stats}. 
Only the first initial condition in the \ace ensemble is used in this analysis. 
Fig. \ref{fig:coarse_tc_tracks}a-c display the spatial distributions of tropical cyclones in the X-SHiELD reference and \ace as well as the \ace track density bias. \ace broadly matches the mean number of tropical cyclones formed per year, averaging 69.0$\pm$6.6 tracks per year to X-SHiELD's 75.3$\pm$6.8. 

When discussing storm intensity in the 100~km data, we use 10~m wind speed as our metric for comparison (instead of IKE which was used to discuss 3~km resolution storms), as the IKE computed using coarse-resolution 10~m wind speeds is very nonlinear with the actual fine-grid IKE.
The histograms of per-track maximum 10~m wind speeds in \ace and 100~km-coarsened X-SHiELD show that the slight low bias in the total number of tropical cyclones in \ace stems from its lower production of weaker cyclones relative to X-SHiELD (Fig. \ref{fig:coarse_tc_tracks}d).
\ace produces fewer tropical cyclones with peak coarse grid 10~m wind speeds below 18 m/s than X-SHiELD (281 in \ace vs. 365 in X-SHiELD) over 10 years. 
The peak wind distribution in \ace is shifted towards stronger storms-- \ace has a slightly higher (by 1 m/s) median peak 10~m wind speed and produces more strong cyclones  with peak wind speed above 18 m/s than X-SHiELD over that same period (409 in \ace vs. 388 in X-SHiELD).

\section{Multivariate \hiroace: 3~km resolution outputs from free-running simulations}
\label{sec:hiroace_results}
Section \ref{sec:hiro_perfect_pred} demonstrates the abilities of the HiROv2 downscaling model to generate realistic distributions of precipitation, winds, and sea level pressure for time and global aggregates as well as snapshots of extreme events. 
Section \ref{sec:ace_coarse_emulator} demonstrates that the 10-year climatologies and tropical cyclone distributions from free-running \ace inference are similar to the X-SHiELD reference run.
Now, we investigate whether combining these two halves of the \hiroace framework produce plausible 3~km resolution predictions of high-impact regional events over a ten-year period. 
We take the first member of the ten-year, 100~km resolution \ace ensemble described in Sec. \ref{subsec:ace_tc_climatology} and use the HiROv2 model to downscale selected time and spatial data ranges.

\begin{figure}[h]
    \centering
    \includegraphics[width=1.0\linewidth]{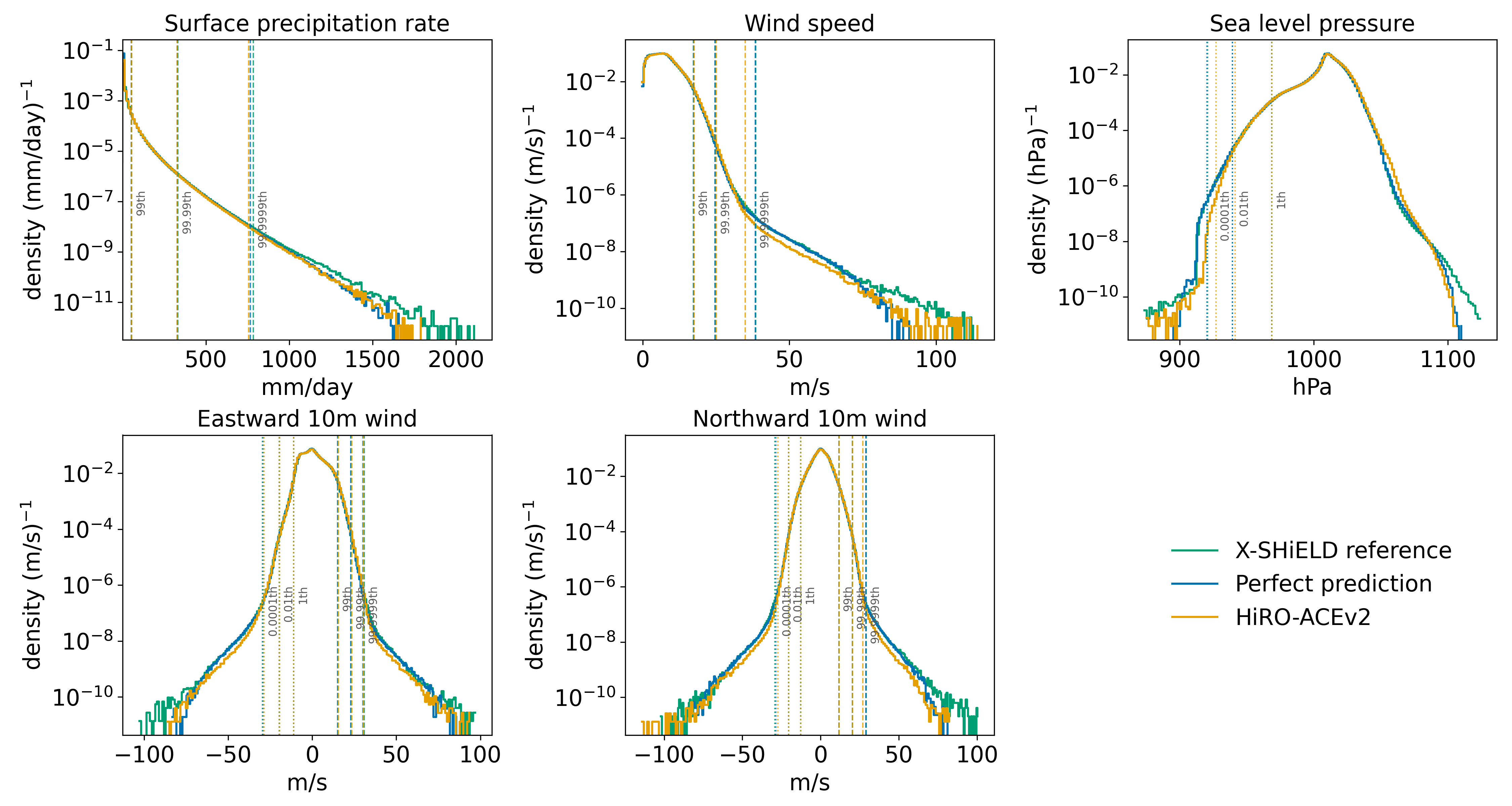}
    \caption{Global (65\degree S--65\degree N) and time aggregrated histograms for one year (2023) of data from the 3~km X-SHiELD reference, 3~km downscaled perfect prediction (a single-sample HiROv2 prediction of the corresponding coarsened 100~km X-SHiELD data), and 3~km free-running \hiroace prediction.}
    \label{fig:hiro_ace_2023_histograms}
\end{figure}

Fig. \ref{fig:hiro_ace_2023_histograms} shows the histograms of downscaled 3~km resolution precipitation, 10~m winds, and sea level pressure fields from 65\degree S--65\degree N over the last year (2023) of the free-running, 10-year \ace inference. The histograms from the \hiroace output are shown alongside the X-SHiELD reference and HiROv2 perfect prediction histograms for the same 2023 year. The perfect prediction distribution is the same as HiROv2 in Fig. \ref{fig:methods_ablation_histograms}. Unlike the perfect prediction case, the downscaling step in the \hiroace results uses coarse-resolution inputs generated by \ace. 
The 3~km \hiroace distributions match the target X-SHiELD distributions out to the extreme tails (e.g., $\sim$99.9999th percentile), indicating that the \ace emulator fine-tuned for this work does a good job in replicating the statistical distributions and spatial character of the coarse-resolution input fields relevant for the downscaling model.
There are slight low biases in \hiroace wind speeds between $\sim35-80$ m/s and sea level pressures below $\sim925$ hPa.

\subsection{Tropical cyclone statistics in free-running \hiroace}
We now compare the intensity distribution of high-resolution tropical cyclones in 3~km resolution \hiroace outputs to the X-SHiELD reference.
We take the same two sets of 100~km resolution tropical cyclone snapshots described in Sec. \ref{subsec:ace_tc_climatology} and compare their corresponding 3~km fields. For the \hiroace cyclones, we generate 8 downscaled samples using the timestep with the highest coarse-grid 10~m wind speed. 
The reference comparison is the set of 3~km resolution reference dataset snapshots corresponding to the 100~km X-SHiELD cyclones.

\begin{figure}[h]
    \centering
    \includegraphics[width=0.65\linewidth]{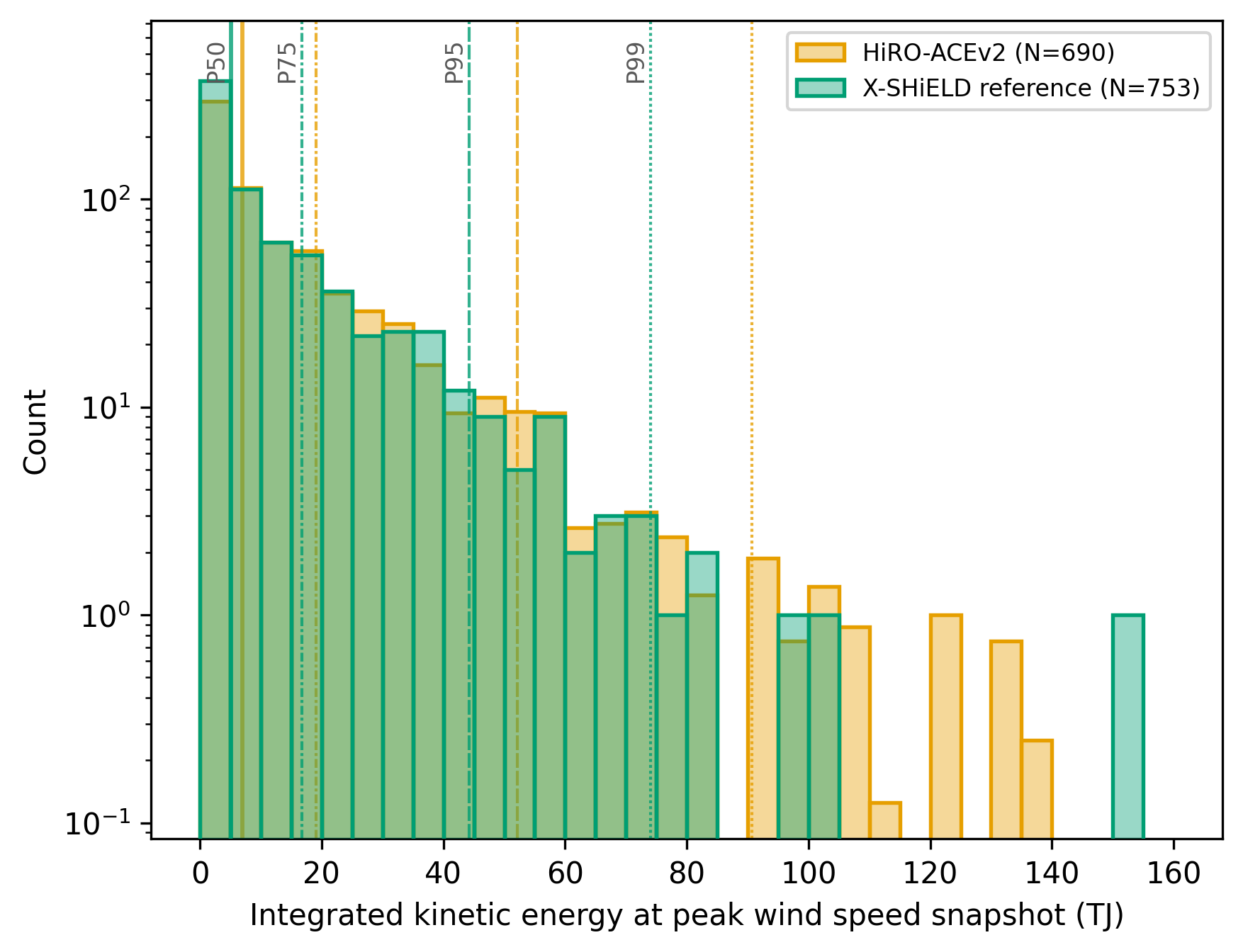}
    \caption{
        Histogram of tropical cyclones' distribution of integrated kinetic energy over 2014-2023 in 3~km resolution \hiroace and X-SHiELD. 
        Vertical lines denote the \{50, 70, 95, 99\}th percentiles of each dataset. 
        The IKE of each cyclone is calculated at the timestep with maximum coarse-field wind speed.
        Each cyclone in \hiroace has 8 generated members, each of which contributes 1/8th count in the histogram.
    }
    \label{fig:hiro_ace_ike_histogram}
\end{figure}

The distribution of tropical cyclone IKE is shown in Fig. \ref{fig:hiro_ace_ike_histogram} for the 3~km \hiroace and X-SHiELD datasets.
Again, we emphasize that the high-resolution cyclones in \hiroace are not downscaled from coarse-resolution X-SHiELD data, so we do not necessarily expect the statistics of tropical cyclones at 3~km resolution to align across the datasets.
Both distributions are similar up to the 75th percentile, beyond which the greater number of coarse-resolution \ace tracks with high wind speeds (as discussed in Sec. \ref{subsec:ace_tc_climatology}) translates into a longer tail of high-IKE tropical cyclones in \hiroace.

\subsection{Case studies: localized, high-impact events in \hiroace}

Here, we plot several case studies of localized events from the downscaled \hiroace data.
Though largely qualitative, these examples demonstrate that the emulation framework naturally gives rise to events with high regional impacts during free-running inference.

\begin{figure}[h]
  \centering
    \includegraphics[width=1.0\linewidth]{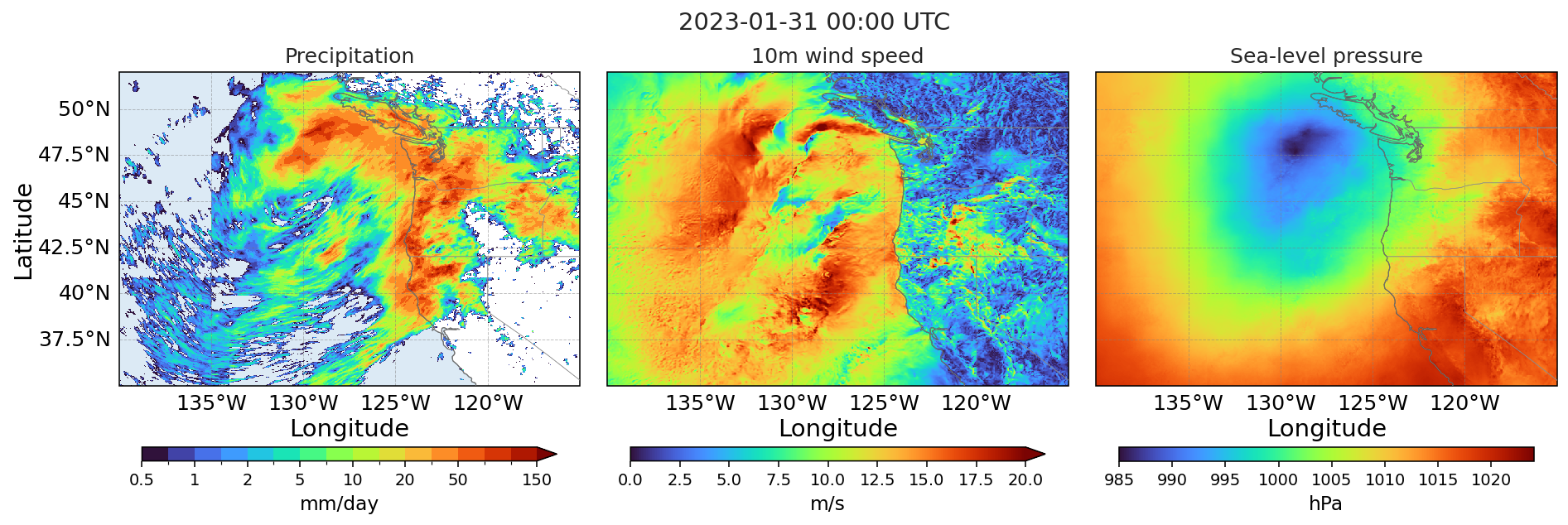}
    \caption{Snapshot of the 6-hourly mean surface precipitation, 10~m wind speed, and sea level pressure fields from an extratropical cyclone and atmospheric river off the western United States and Canada in \hiroace.}
    \label{fig:hiro_ace_event_ar}
\end{figure}

Fig. \ref{fig:hiro_ace_event_ar} shows the precipitation, wind speed, and sea level pressure fields from an extra-tropical cyclone off the western coast of Vancouver Island with an atmospheric river feeding in from the southwest.
The simulated 3~km-resolution precipitation field displays realistic texture over ocean, banding around the cyclone, and strengthening over topography as the system moves over land.

\begin{figure}
    \centering
    \includegraphics[width=1.0\linewidth]{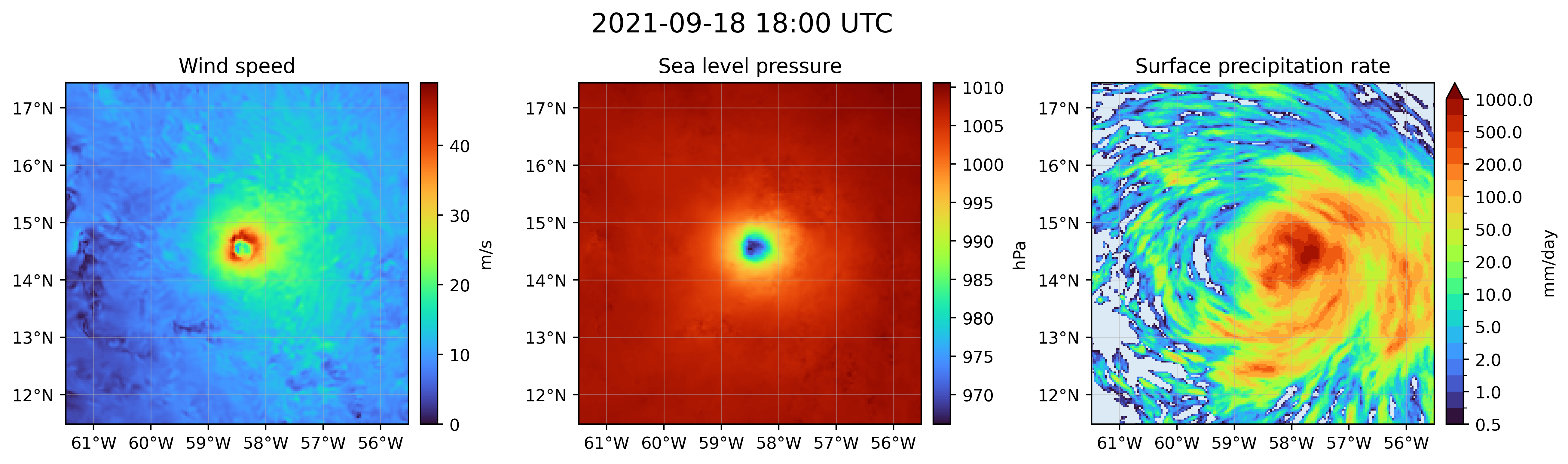}
    \caption{10~m wind speed, sea level pressure, and precipitation fields from a North Atlantic hurricane in free-running \hiroace.}
    \label{fig:hiro_ace_tc}
\end{figure}

Fig. \ref{fig:hiro_ace_tc} shows a strong North Atlantic hurricane from the \hiroace run.
The 10~m wind speed, sea level pressure, and precipitation fields are all qualitatively similar in spatial structure to the perfect prediction comparison in Fig. \ref{fig:perfect_pred_Phl_tc}.
We also verify that the full set of downscaled tropical cyclones in \hiroace have similar morphology as in the X-SHiELD reference (\ref{app:hiro_ace_tc_shape}).

\begin{figure}
    \centering
    \includegraphics[width=1.0\linewidth]{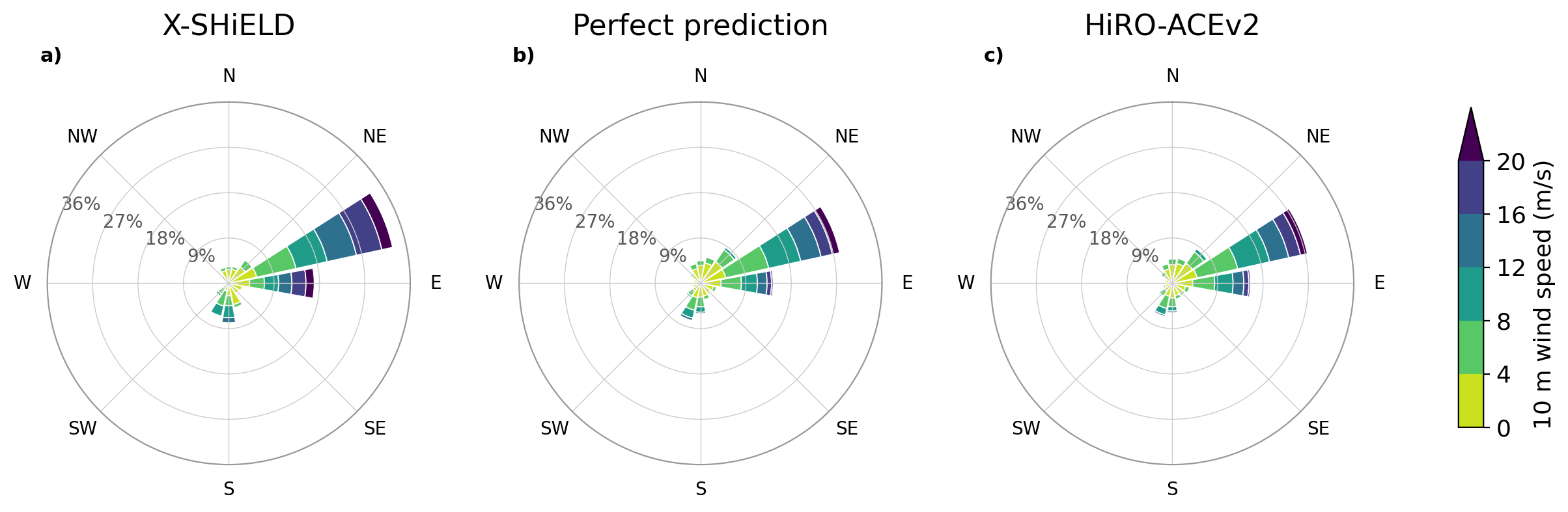}
    \caption{Wind rose diagrams for 10~m wind aggregated over 2014-2023 3~km a) X-SHiELD, b) perfect prediction HiROv2 downscaled, and c) \hiroace datasets over Senj, Croatia.}
    \label{fig:hiro_ace_wind_rose}
\end{figure}

Finally, as a demonstration of how \hiroace can be used to efficiently generate high-resolution, localized predictions over multiyear timescales, we plot wind rose climatologies for Senj, Croatia over the 2014-2023 period (Fig. \ref{fig:hiro_ace_wind_rose}). The coastal town of Senj is well-known for extreme wind conditions that arise during Bora events.  Both the HiROv2 perfect prediction and free-running \hiroace climatologies are compared to 3~km X-SHiELD.
Qualitatively, the free-running \hiroace simulation predicts a similar wind climatology to X-SHiELD at this location; the dominant feature across all three columns is the frequent occurrence of Bora winds from the E and ENE directions. 
However, both the HiROv2 perfect prediction and \hiroace climatologies have almost the exact same low bias ($\sim15\%$) relative to X-SHiELD along the E and ENE wind directions-- for this particular case study, the bias observed in \hiroace is due to errors in the HiROv2 downscaling of the 10~m winds, and not from coarse-resolution biases in \ace inference.

\subsection{Computational cost}

HiRO-ACE inference computation is dominated by the downscaling step: the coarse \ace emulation can generate a year of data in under 5 minutes, while downscaling that year of global data takes $\sim$2.75 days on a single B200 GPU.
Overall, the inference energy cost to produce a yearlong, global dataset at 3~km resolution via HiRO-ACEv2 is roughly 0.3\% of the cost to produce a year of data by running the X-SHiELD model.
This value is slightly lower than in \perkins{} due to HiROv2 code optimizations introduced between versions which speed up training and inference.
The wall clock time to generate a full global 3~km dataset via HiRO-ACEv2 (on 8 NVIDIA B200 GPUs) is $\sim$33$\times$ faster than via the X-SHiELD simulation (on 27648 cores on AMD EPYC 9654 processors); this could be further accelerated by increasing the HiROv2 prediction parallelization over more GPUs.

Since the spatial region and time range selected for downscaling from the emulator outputs can be set by the user, computational cost is further reduced by up to a factor of up to $\sim$250 by downscaling a smaller number of patches of the minimum output size 16$^\circ\times$16\degree-- this is large enough to cover the spatial extent of the extreme events studied in this work.
A year of data for a single coarse-resolution patch of this size downscales in 12 (15, 30) minutes on a single NVIDIA B200 (H100, L40S) GPU.
We are also actively exploring distillation methods which could increase inference speeds by over an order of magnitude in future HiRO models while preserving prediction skill \citep{fastgen2026}.

\section{Conclusions}

This work introduces \hiroace, a system for emulating the precipitation, near-surface winds, and sea level pressure outputs from the 3~km resolution X-SHiELD model.
Building upon our previously released model from \perkins, we train new versions of both \hiroace components: the 100~km coarse-resolution GCM emulator (\ace) and the diffusion-based downscaling model (HiROv2).
Simultaneous prediction of multiple high-resolution output fields, newly introduced here in HiROv2, allows the model to downscale a wider range of extreme events than the precipitation-only HiROv1.

We focus on capturing the statistics of extreme events when training HiROv2.
Several changes to the downscaling diffusion model substantially improve its generation of high-percentile outputs: separate denoisers for high and low noise ranges, and modified noise sampling, data sampling, and noise-weighted loss during training. 
After these training updates, HiROv2 matches the global and annual aggregated target output distributions to within 3\% of the reference 99.9999th percentiles for all output fields.

When downscaling 100~km resolution X-SHiELD inputs, HiROv2 closely reproduces the aggregated target 3~km X-SHiELD statistics and generates downscaled snapshots of extreme events that are consistent with the characteristics of the reference.
It predicts the integrated kinetic energy of high-resolution tropical cyclones well, with consistently low RMSE ($<$1 TJ) and no significant bias across tropical cyclones in the held out set.
For the most part, the downscaled tropical cyclones exhibit realistic structures in their sea level pressure, 10~m wind, and precipitation fields, though the predicted wind and pressure fields are somewhat more asymmetric in the inner eye than the reference storms.
We also demonstrate that HiROv2 resolves strong near-surface winds over topography with realistic downslope flows.

An updated version of the 100~km \ace emulator of coarsened X-SHiELD is trained for use in \hiroace. 
Over a 10 year inference period, \ace's tropical cyclone climatology as well as its climatological fields and aggregated histograms of precipitation, 10~m winds, and sea level pressure are in close agreement with the X-SHiELD reference.
These results show that \ace coarse-resolution simulations are suitable for generating long-term high-resolution climate datasets.

A full year of global 3~km data is generated by combining the coarse-resolution \ace emulator with the HiROv2 downscaling model. 
Time-aggregated histograms for the \hiroace simulation data align closely with the 3~km X-SHiELD reference, despite the fact that the downscaling is performed on an unpaired free-running simulation. 
Over a 10 year free-running rollout, the integrated kinetic energy distribution of tropical cyclones produced in \hiroace is also comparable to that of the same 10 year period in X-SHiELD.
In addition to tropical cyclones, case studies of various event snapshots downscaled from the full-length 10 year rollout show that \hiroace also generates a range of extreme events similar to those found in X-SHiELD.

The results presented here support the case for \hiroace as a flexible and computationally efficient framework for researchers to stochastically generate ensembles of high resolution precipitation, near-surface winds, and sea level pressure fields for extreme events over long ($\geq10$ year) simulations.
The computational costs of running \hiroace to produce global high-resolution data are roughly 300x lower than those of the physics-based X-SHiELD model.
For use cases that only require regional downscaling, these computational costs can be further reduced by over two orders of magnitude, depending on the domain size.
Work is in progress to distill the trained HiROv2 downscaling model, which could further increase this efficiency gain by over an order of magnitude.

The downscaling HiROv2 model is trained on X-SHiELD and thus can only be paired with the X-SHiELD-tuned version of \ace; in its present form \hiroace cannot be used to generate high-resolution fields consistent with other climates, such as a warmed climate or ERA5 reanalysis. In upcoming work, we will utilize new X-SHiELD datasets (e.g. X-SHiELD nudged to ERA5, or X-SHiELD in a +4~K warmed climate) to allow the model to operate in different climates.
As the current HiRO-ACE framework currently treats timesteps as independent generations, future work will aim to improve temporal consistency between consecutive downscaled timesteps.
We also plan to add more output channels to future versions of HiRO, such as near-surface air temperature or frozen precipitation, which would enable an even wider range of use cases.

%
%

\section*{Open Research Section}
Code supporting training, inference and evaluation for HiRO-ACE can be found at \url{https://github.com/ai2cm/ace/tree/v2026.8.1} \citep{ACEZenodo}. Code documenting the exact training configurations and scripts to generate figures can be found at \url{https://github.com/ai2cm/hiro-acev2-manuscript} and \url{https://zenodo.org/records/23325424}. Model weights, forcing data, initial conditions, and configurations to generate example HiRO-ACE outputs over the 2014--2023 period are available at \url{https://huggingface.co/allenai/HiRO-ACEv2}. The C96 SHiELD dataset that was used for pretraining ACE2S-X-SHiELD is available as analysis-ready cloud-optimized zarr stores (15.8 TB total) at \url{https://huggingface.co/buckets/allenai/ACE2S-SHiELD-plus-reference-data}. The X-SHiELD data used for fine-tuning ACE2S-X-SHiELD and training HiRO are available at \url{https://huggingface.co/buckets/allenai/ai2cm-xshield}.

\section*{Conflict of Interest declaration}
The authors declare there are no conflicts of interest for this manuscript.

\acknowledgments
Ai2 is supported by the Allen Fund for Science and Technology. We thank the NOAA Geophysical Fluid Dynamics Laboratory for performing the X-SHiELD simulation and supporting our use of its output. We acknowledge helpful discussions with Akshay Subramanian from NVIDIA and Jan-Huey Chen and Baoqiang Xiang from NOAA GFDL. AI tools (Claude Sonnet 4.6-5.5 Claude Opus 4.6–5, Cursor Composer 1 and 2) were used in this work to generate code, edit configuration files, generate plotting and analysis scripts, and proofread the manuscript. 

%
%

\bibliography{references}

\appendix

\section{Comparisons to HiRO-ACEv1}
\label{app:v1_comparison}
Table \ref{tab:hiro-v1-v2} summarizes the main differences between HiRO-ACE v1 \citep{perkins2026} and v2 (this work).

\begin{table}[ht]
\centering
\small
\renewcommand{\arraystretch}{1.25}
\begin{tabularx}{\textwidth}{>{\raggedright\arraybackslash}p{3.2cm} >{\raggedright\arraybackslash}X >{\raggedright\arraybackslash}X}
\toprule
\textbf{~} & \textbf{v1} & \textbf{v2} \\
\midrule
\textbf{Input variables} &
Coarse surface precipitation rate (6 hr average), 10\,m eastward and northward winds, high-res static surface height &
Coarse surface precipitation rate (6 hr average), 10\,m eastward and northward winds, sea level pressure, high-res static surface height and land surface fraction \\
\midrule
\textbf{Output variables} &
High-res surface precipitation rate (6 hr average) &
High-res surface precipitation rate (6 hr average), 10\,m eastward and northward winds, sea level pressure \\
\midrule
\multicolumn{3}{l}{\textbf{HiRO diffusion model training}} \\
\quad Noise sampling &
Log normal distribution with $p_{\mathrm{mean}} = -1.2$, $p_{\mathrm{std}} = 1.8$ &
Log uniform distribution with $(p_{\mathrm{min}}, p_{\mathrm{max}}) = (0.005, 200)$ and $(200, 2000)$ for the low- and high-range denoisers respectively \\
\quad Num.\ parameters &
30M &
32M (low-range denoiser), 46M (high-range denoiser) \\
\quad Num.\ denoisers &
Single denoiser &
Two denoisers, for the $(0.005, 200)$ and $(200, 2000)$ ranges of the noise schedule. The low-range denoiser has updated noise weighting and downweights precipitation in the loss. \\
\quad Data sampling &
Uniform weighting over $-66^\circ$ to $70^\circ$ latitude &
$8\times$ oversampling of patches with center latitude in $(-30^\circ, 30^\circ)$ \\
\midrule
\multicolumn{3}{l}{\textbf{ACE2S-X-SHiELD training}} \\
\quad Dataset &
Base checkpoint pretrained on ERA5 &
Base checkpoint pretrained on ensembled SHiELD data \\
\quad Fine-tuning &
Fine-tuned on 100\,km X-SHiELD &
Fine-tuned on the same SHiELD dataset as the base model with sea level pressure added as a diagnostic, then fine-tuned on 100\,km X-SHiELD \\
\bottomrule
\end{tabularx}
\caption{Comparison of HiRO-ACE v1 and v2.}
\label{tab:hiro-v1-v2}
\end{table}

\section{HiROv2 details}
\label{app:hiro_details}

\subsection{Training details}
\label{app:hiro_training}
Our downscaling model HiROv2 is conditioned on i) the coarse resolution X-SHiELD fields corresponding to 6 hourly averaged precipitation, instantaneous 10~m eastward and northward winds, and instantaneous sea level pressure, interpolated bicubically to the fine-resolution X-SHiELD grid and ii) the static land fraction and surface height field from X-SHiELD. 

\begin{table}[t]
\centering
\caption{Training configurations for the HiROv2 diffusion downscaling denosing models.
Where two values are given, they refer to the low-$\sigma$ and high-$\sigma$
denoiser respectively; all other settings are shared.}
\label{tab:train_config}
\begin{tabular}{l p{8cm}}
\toprule
\textbf{Hyperparameter} & \textbf{Value (low-$\sigma$ / high-$\sigma$)} \\
\midrule
\multicolumn{2}{l}{\textit{Architecture}} \\
Backbone & Song-style U-Net \citep{physicsnemo}, modified to omit attention at lowest resolution decoder block  \\
Base channels & 128 \\
Channel multipliers & $[1,2,2,2,1,1]$ / $[1,2,2,2,2,2]$ \\
Embedding channel multiplier & 6 \\
Residual blocks per resolution & 1 \\
Attention resolutions & none \\
\midrule
\multicolumn{2}{l}{\textit{Diffusion noise sampling}} \\
$\sigma_{\min}$ & $5\times10^{-3}$ / $200$ \\
$\sigma_{\max}$ & $200$ / $2000$ \\
Training noise distribution & log-uniform over $[\sigma_{\min},\sigma_{\max}]$ \\
\midrule
\multicolumn{2}{l}{\textit{Objective}} \\
Loss & MSE \\
Noise weighting exponent & 0.75 / 1.0 \\
Channel weight (\texttt{PRATEsfc}) & 0.2 / 1.0 (1.0 for all other channels) \\
\midrule
\multicolumn{2}{l}{\textit{Optimization}} \\
Optimizer & Adam \\
Learning rate & $1\times10^{-4}$ \\
EMA decay & 0.999 \\
Precision & mixed, bf16 \\
\bottomrule
\end{tabular}
\end{table}

\begin{figure}[h]
    \centering
    \includegraphics[width=0.65\linewidth]{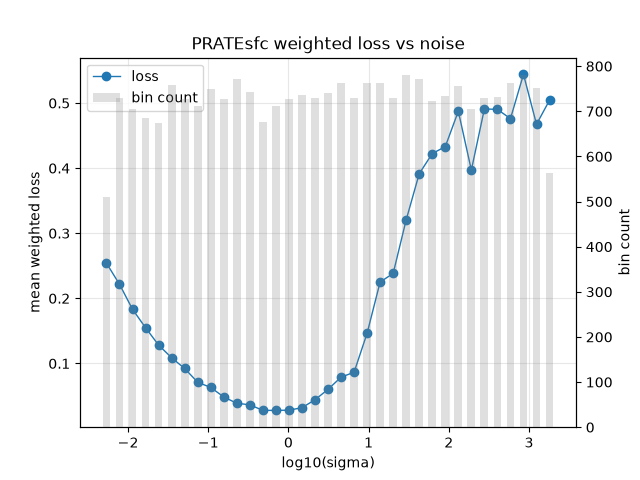}
    \caption{Weighted loss vs. noise for the precipitation channel during a random epoch from a model training run across the full noise schedule range (0.005, 2000).}
    \label{fig_sup:noise_vs_loss}
\end{figure}

The training noise log uniform distribution upper bound value $\sigma_{max}$ was chosen by plotting the binned weighted loss (c.f. \citet{karras2022} for description of weighting) versus noise, and observing the approximate noise value at which the right hand side of the curve started to plateau. This indicates the noise values at which the largest values in the training data are fully swamped by noise. An example of the weighted loss versus noise curve for precipitation is shown in Fig. \ref{fig_sup:noise_vs_loss}. The value $\sigma_{max}=2000$ used as the upper bound of the log uniform distribution during the high-noise denoiser training, as well as the starting value in the inference noise schedule, was picked to be safely after the curve had plateaued.

Training data is selected from latitudes between -66$^\circ$ and 70$^\circ$ in order to avoid potential issues due to grid spacing near the poles.
For each batch, the near-global domain is divided into 16$^{\circ}\times$16$^\circ$ patches. A random offset in both latitude and longitude between 0-16\degree is chosen for each batch, such that the training patches are sampled from a randomly shifted grid; this allows inference to perform well on arbitrary patch domains. When oversampling tropical regions, the same number of patches from the full shifted domain is sampled with replacement from a distribution defined by patches' center coordinates, where patches within $\pm$30\degree are upweighted by 8x.

The best checkpoint for each denoiser is selected as the one with the lowest continuous ranked probability score (CRPS) on the validation set. 
Validation was performed every 20 epochs on a subset of 162 snapshots in the West Pacific region bounded by latitudes (0\degree, 35\degree) and longitudes (75\degree, 195\degree) to reduce time spent on computing validation outputs, as this step can be a bottleneck during training.
Note that this selection criteria is different from the checkpoint selection procedure in \perkins, which used an absolute normalized bias metric to select the final checkpoint. 
We found that after applying the training updates described in Section \ref{subsec:model_training}, checkpoints' skill in predicting extreme tails became more stable over training and CRPS was equivalently skillful at capturing this while also being a more standard metric.

The lower sigma range diffusion model is trained for 105,600 gradient steps. The higher sigma range diffusion model is trained for 52,800 steps. Both trainings use a batch size of 48 timesteps.
Full training parameters are given in Table \ref{tab:train_config}.

\subsection{Inference details}
\label{app:hiro_inference}

For downscaling inference, we use the standard EDM algorithm with 18 denoising steps, $\sigma_{min}=0.005,~\sigma_{max}=2000,~\rho=7$, and $S_{churn}=0$.
For each value of $\sigma$ in the denoising schedule, the appropriate low- or high-$\sigma$ range denoiser is called. The thresholds that determine which denoiser is routed to are the same as their training noise bounds in Table \ref{tab:train_config}.

The model is required to predict downscaled regions of 16$^{\circ}\times$16$^\circ$ or larger in either dimension, to match its training patch size. 
While the UNet architecture is capable of outputting single generated fields with different dimensions from its training set, in practice we find that this can lead to larger errors and thus require the model to input and output 16$^{\circ}\times$16$^\circ$ fields.
If the requested inference domain size is larger than 16$^{\circ}\times$16$^\circ$, the model will generate the set of output patches required to cover the domain with a 1\degree overlap between output patches. Each patch is independently generated and the set is composited together by averaging within the overlapping edges.
In the future, spatial coherence over large output regions could be improved by using a multidiffusion strategy where a local denoiser operates over patches and then aggregates over the domain at each diffusion denoising step \citep{bartal2023}.

\section{Perfect prediction: Additional results}

\begin{sidewaysfigure}
    \centering
    \includegraphics[width=1.0\linewidth]{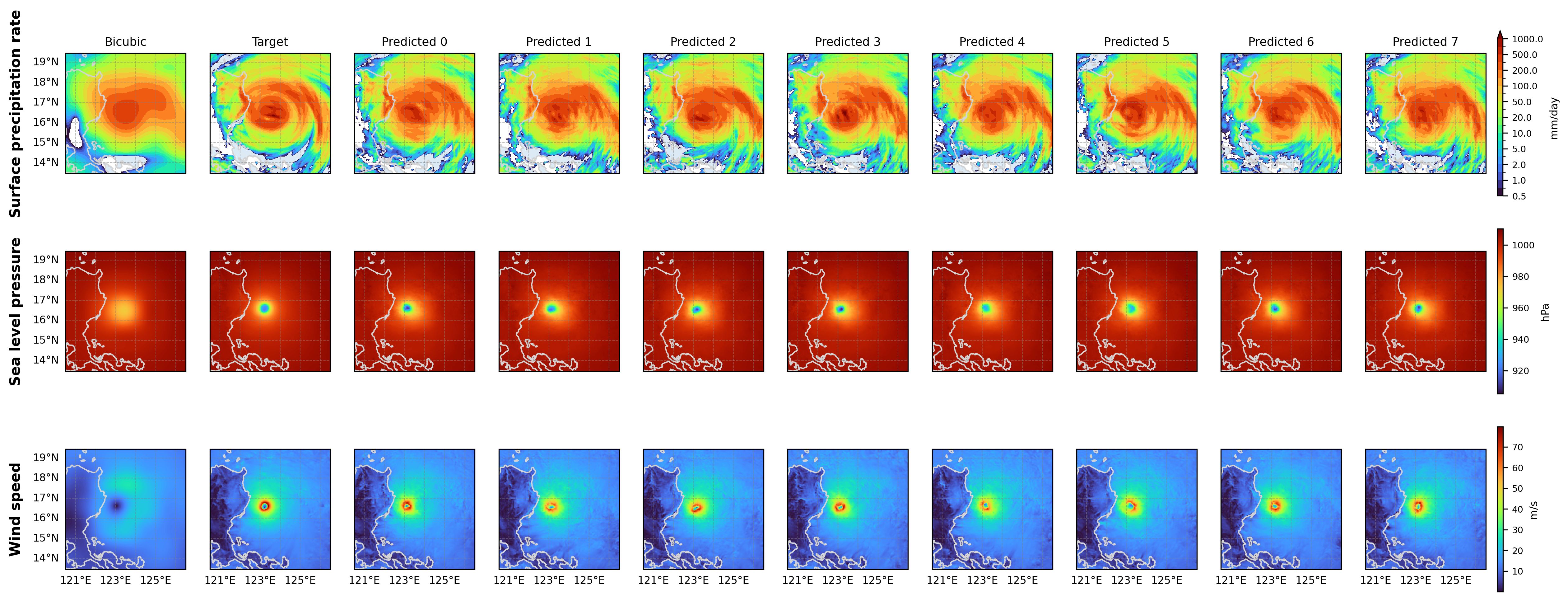}
    \caption{Full 8 member ensemble for the tropical cyclone event shown in Fig. \ref{fig:perfect_pred_Phl_tc}.}
    \label{fig_sup:perfect_pred_Phl_tc_ensemble}
\end{sidewaysfigure}

Fig. \ref{fig_sup:perfect_pred_Phl_tc_ensemble} shows the full HiRO-downscaled 8 member ensemble of the tropical cyclone event shown in Fig. \ref{fig:perfect_pred_Phl_tc}.
The qualitative findings described in that example are all broadly applicable across the generated ensemble.

\begin{figure}
    \centering
    \includegraphics[width=1.0\linewidth]{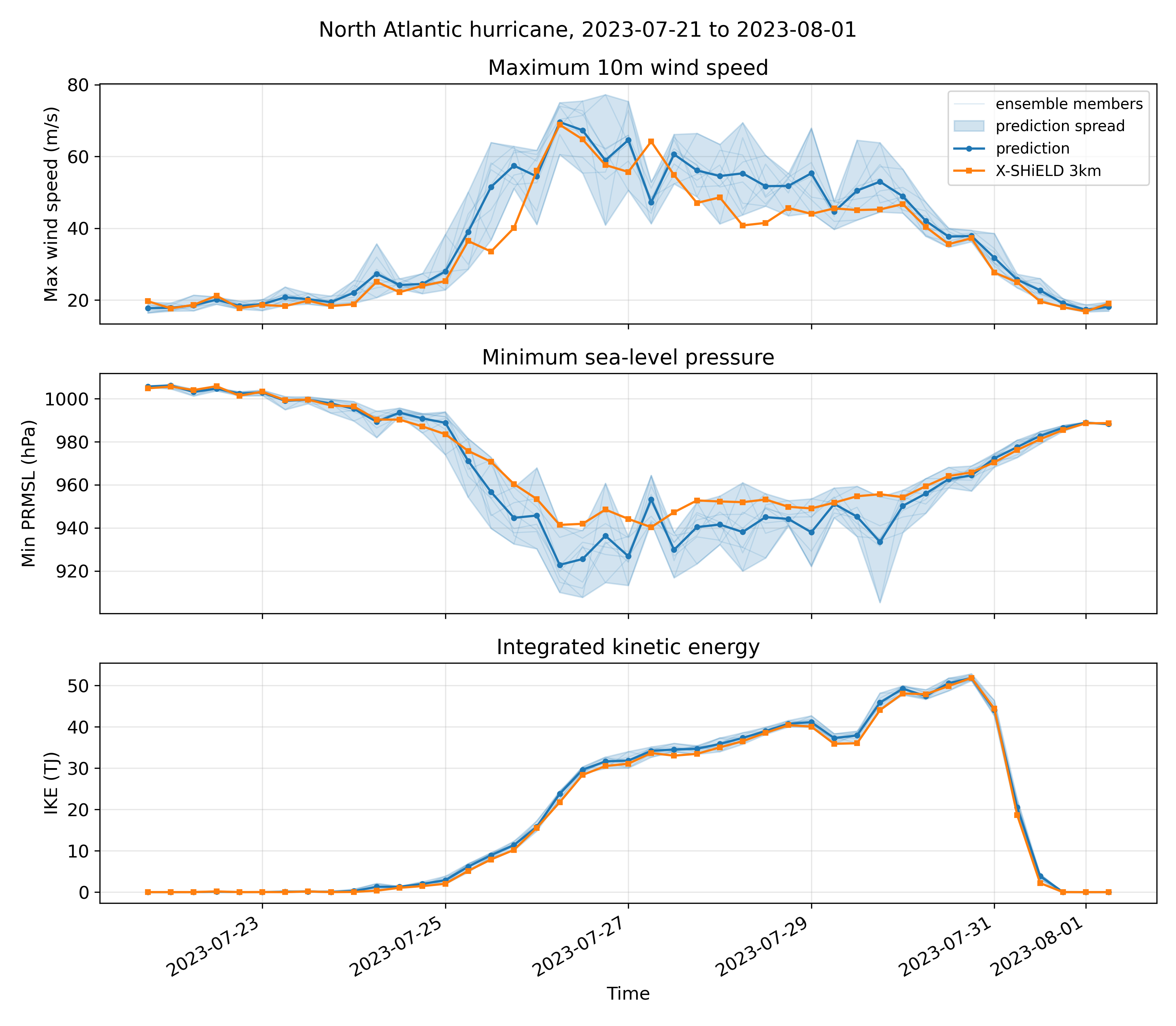}
    \caption{Example of a time series of downscaled snapshots of a strong North Atlantic hurricane. The eight individual ensemble members are shown as faint blue lines. The thick blue line shows the ensemble mean, and the shaded band indicates ensemble spread.}
    \label{fig_sup:perfect_pred_tc_time_series}
\end{figure}

Fig. \ref{fig_sup:perfect_pred_tc_time_series} shows example time series of the maximum 10~m wind speed, minimum sea level pressure, and IKE in downscaled and reference 3~km-resolution fields for a strong North Atlantic hurricane. 
The lack of temporal coherence across consecutive time steps is evident in the maximum wind speed and minimum sea level pressure time series for individual ensemble members. This is to be expected, as each time step is downscaled independently.
Taking the ensemble mean smooths out a large amount of this variance.
HiROv2's tendency to overshoot the sharp minimum sea level pressure cutoff in tropical cyclone eyes, as discussed in Section \ref{subsec:perfect_pred_tc_example}, is also apparent throughout the time series of minimum sea level pressure.

\section{Coarse-resolution \ace emulator training details}
\label{app:ace-training}
The full training procedure for the 100~km \ace emulator is as follows. 
We start with a checkpoint associated with the ACE2S-SHiELD+ model from \citet{clark2026}, which is trained on the dataset described in Section~\ref{sec:ace2s-shield-plus-dataset}. 
It should be noted that our starting checkpoint differs from the released checkpoint, which was selected using the best inline inference error \citep{shieldplusckpt2026}. We use the checkpoint with the best validation loss from the final multi-step tuning stage as our base.

We then tune this base checkpoint on the same dataset, but with sea level pressure added as an additional diagnostic output, since the ACE2S-SHiELD+ model did not include this as an output.
Additional diagnostic outputs can be easily added to an existing ACE checkpoint by loading a pretrained checkpoint's weights and appending the new variable name(s) to the end of the diagnostic outputs list. We do not freeze any weights.
We use the same multi-step training configuration as \citet{clark2026} with the exception of disabling the energy corrector. 

The resulting checkpoint is then fine-tuned once more on the ten years of coarsened X-SHiELD data.
The training configuration in this final tuning step is the same as in \citet{clark2026}, except that the energy corrector is again disabled, and the loss weight on precipitation is increased from 0.5 to 5.

The base checkpoint, which was produced for \citet{clark2026}, trained over $\sim$~490,000 gradient descent steps. Tuning to include sea level pressure took $\sim$~70,000 steps, and the final fine-tuning step on X-SHiELD took $\sim$~73,000 steps.

\section{High-resolution tropical cyclone morphology in free-running \hiroace}
\label{app:hiro_ace_tc_shape}

\begin{sidewaysfigure}
    \centering
    \includegraphics[width=1.0\linewidth]{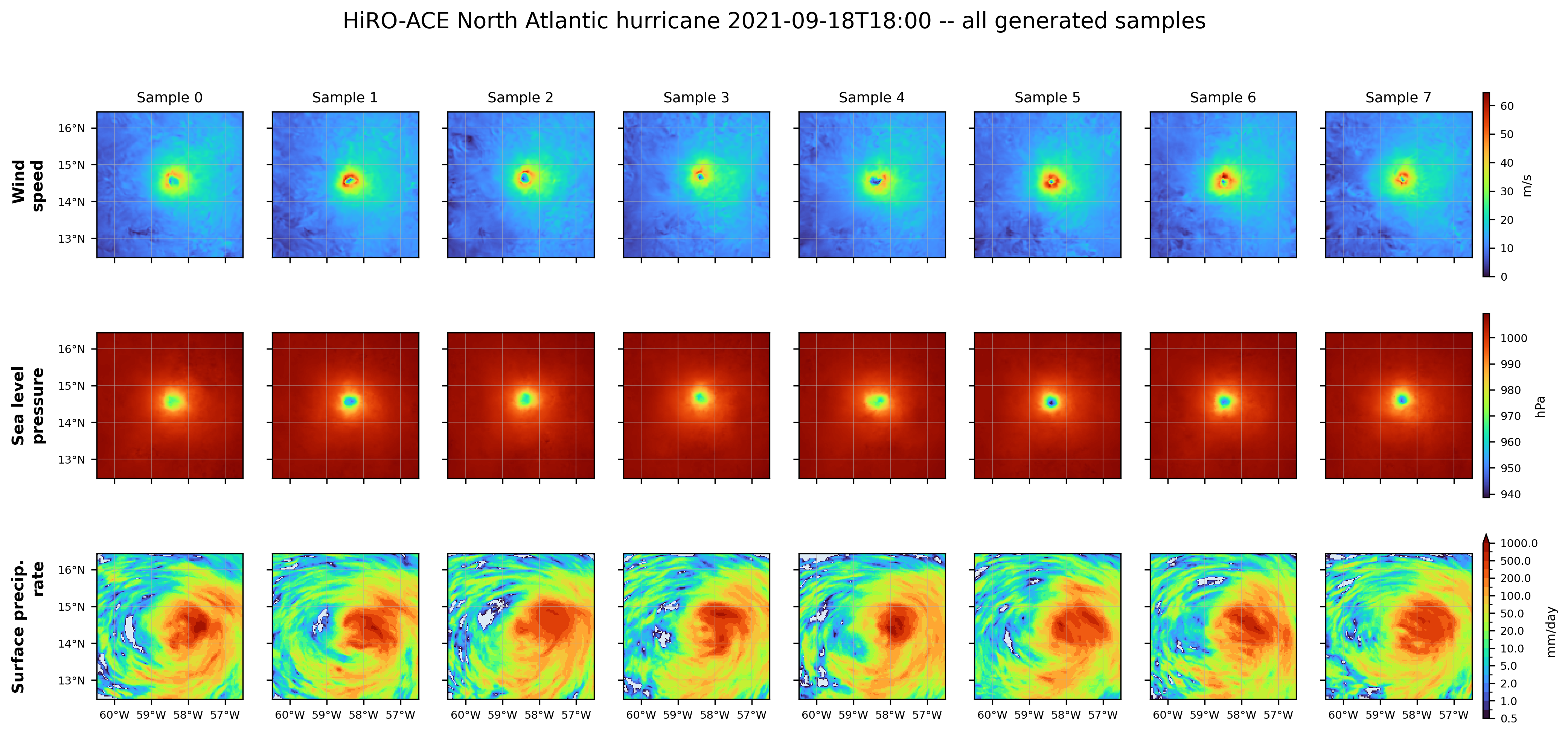}
    \caption{Example of a 3~km resolution tropical cyclone in the free-running \hiroace ten year inference run.}
    \label{fig_sup:hiro_ace_tc_example_ensemble}
\end{sidewaysfigure}

The full 3~km resolution generated ensemble corresponding to the \hiroace North Atlantic hurricane shown in Fig. \ref{fig:hiro_ace_tc} is shown in Fig. \ref{fig_sup:hiro_ace_tc_example_ensemble}.

\begin{figure}[h]
    \centering
    \includegraphics[width=1.0\linewidth]{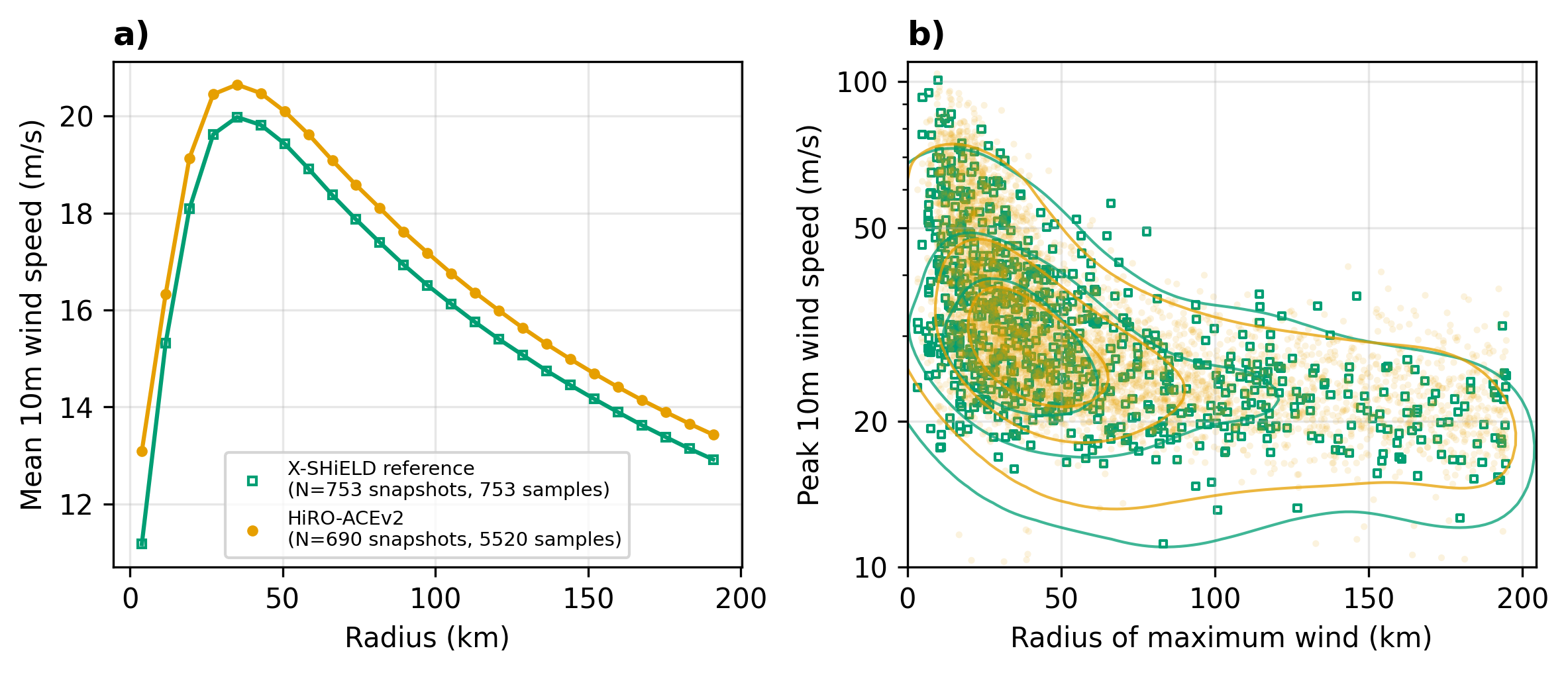}
    \caption{a) Mean radial wind profile of 3~km resolution tropical cyclones over 2014-2023 in the X-SHiELD reference dataset (green) and the free-running \hiroace output (orange). b) Scatter plot and density contours showing the relation between maximum wind speed and radius of maximum wind speed across the set of tropical cyclones in each dataset.}
    \label{fig_sup:radial_wind_hiro_ace}
\end{figure}

Fig. \ref{fig_sup:radial_wind_hiro_ace}a shows the mean radial 10~m wind profiles of the \hiroace and X-SHiELD cyclones. 
For the most part, the radial wind profiles of the \hiroace and X-SHiELD tropical cyclones are remarkably similar. The scatterplot of each set of tropical cyclones' radius of maximum wind versus peak 10~m wind speed is shown in Fig. \ref{fig_sup:radial_wind_hiro_ace}b to illustrate that the full distribution of shapes also roughly agrees between \hiroace and X-SHiELD.
The mean 10~m wind speed in \hiroace is slightly higher ($\lesssim$1 m/s at peak) than the reference mean, likely due to the coarse-resolution \ace emulator's aforementioned tendency to produce tropical cyclones with higher peak wind speeds. 
The centermost few bins of the \hiroace radial wind profile also have slightly higher wind speeds relative to the peak than in X-SHiELD; this is consistent with the asymmetry observed in the downscaled wind outputs which also inflates the central wind speeds in the perfect prediction case.

\end{document}